\documentclass[aps,prd,
preprint,
floatfix,nofootinbib,groupedaddress,showpacs]{revtex4}
\usepackage{graphicx}
\usepackage{amsmath}
\usepackage{latexsym}
\usepackage{here}
\usepackage{array}
\usepackage{dcolumn}
\usepackage{bm}
\usepackage{epsfig}

\usepackage[dvips]{color}
\definecolor{Black}{named}{Black}
\definecolor{Red}{named}{Red}
\definecolor{Blue}{named}{Blue}

\newcommand{\dd}{\mbox{{\rm d}}}

\newcommand{\Lumint}{{\cal L}_{\rm int}}

\def\epem{\ifmmode e^+e^-\else $e^+e^-$\fi}
\def\to{\rightarrow}

\def\mpl{\ifmmode \overline M_{Pl}\else $\bar M_{Pl}$\fi}
\def\beq{\begin{equation}}
\def\be{\begin{equation}}
\def\beqn{\begin{eqnarray}}
\def\ee{\end{equation}}
\def\eeq{\end{equation}}
\def\eeqn{\end{eqnarray}}

\begin{document}
\vspace*{-20mm}
DESY 11-139
\vspace*{2mm}

\title{Spin-analysis of s-channel diphoton resonances at the LHC}

\author{
M.\ C.\ Kumar$^{a,}$\footnote{mckumar@hri.res.in \\
{\em Present address}: Deutsches Elektronensynchrotron DESY, Platanenallee 6,
D-15738 Zeuthen, Germany}\hspace{.4cm}
Prakash Mathews$^{b,}$\footnote{prakash.mathews@saha.ac.in}\hspace{.4cm}
A. A. Pankov$^{c,}$\footnote{pankov@ictp.it}\hspace{.4cm}
N. Paver$^{d,}$\footnote{nello.paver@ts.infn.it}\hspace{.4cm}
V.\ Ravindran$^{a,}$\footnote{ravindra@hri.res.in} \hspace{.4cm}
A. V. Tsytrinov$^{c,}$\footnote{tsytrin@rambler.ru}}
\vspace{.5cm}

\affiliation{
 $^{a}$ Regional Centre for Accelerator-based Particle Physics\\
Harish-Chandra Research Institute, Chhatnag Road, Jhunsi, Allahabad 211 019, India\\
 $^{b}$ Saha Institute of Nuclear Physics, 1/AF Bidhan Nagar, Kolkata 700064, India\\
 $^{c}$ The Abdus Salam ICTP Affiliated Centre, Technical University of Gomel, 246746 Gomel, Belarus\\
 $^{d}$ University of Trieste and INFN-Trieste Section, 34100 Trieste, Italy
}


\begin{abstract}

The high mass neutral quantum states envisaged by theories of
physics beyond the standard model can at the hadron colliders
reveal themselves through their decay into a pair of photons. Once
such a peak in the diphoton invariant mass distribution is
discovered, the determination of its spin through the distinctive
photon angular distributions is needed in order to identify the
associated nonstandard dynamics. We here discuss the
discrimination of the spin-2 Randall-Sundrum graviton excitation
against the hypothesis of a spin-0 exchange giving the same number
of events under the peak, by means of the angular analysis applied
to resonant diphoton events expected to be observed at the
LHC.
The spin-0 hypothesis is modelled by an effective interaction of a
high mass gauge singlet scalar particle interacting with the
standard model fields. The basic observable of our analysis is the
symmetrically integrated angular asymmetry $A_{\rm CE}$,
calculated for both graviton and scalar $s$-channel
exchanges to next-to-leading order in QCD.
\end{abstract}
\maketitle

\section{Introduction}
Diphoton final states represent a very important testing ground
for  the standard model (SM), for example they may be one of the
main discovery channels for the Higgs boson searches at the
CERN LHC.
Moreover, similar to the case of dileptons, the inclusive
production of two-photon high mass resonance states at the LHC:
\begin{equation}
p+p\to \gamma\gamma+X,
\label{proc}
\end{equation}
is considered as a powerful, clean test of New Physics (NP), would
an excess of $\gamma \gamma$ events be observed with respect to
the prediction from the SM cross section.
\par
One NP scenario of particular importance is the case of the spin-2
Kaluza-Klein (KK) graviton excitations predicted by the Randall-Sundrum
(RS) model of gravity in one warped spatial extra
dimension~\cite{Randall:1999ee}. This model suggests a rich
phenomenology that includes the production of diphoton resonances,
to be explored at collider energies, see, for example,
Refs.~\cite{Davoudiasl:2000wi,Davoudiasl:2008qm, Kisselev:2008xv}.
The existence of such graviton excitations can be signalled by the occurrence of peaks in the invariant mass distribution of the photon pairs and, indeed, the lowest lying predicted diphoton peak has recently been searched for
in experiments at the $p ~{\bar p}$ Fermilab Tevatron collider
\cite{Aaltonen:2010cf,Abazov:2010xh}, and at the 7 TeV $p ~ p$ LHC collider with time-integrated luminosity of the order of
40 ${\rm pb}^{-1}$ \cite{atlas1,cms1}. In these experiments, exclusion mass limits on
the lightest RS resonance of the TeV order have been set, and
graviton mass scales larger than 1 TeV will certainly be in the
kinematical reach of LHC.
\par
Assuming that a diphoton peak at an invariant mass value $M_R$ is
observed, its association to a specific NP scenario would be
possible only if we are able to discard other competitor models,
potential sources of the peak itself with same $M_R$ and same
number of events. Basically, for any nonstandard model one can
define, on the basis of the foreseeable statistics and
uncertainties, a {\it discovery reach} on the relevant heavy
resonance $R$ as the upper limit of the range in $M_R$ where, in a
specific domain of the model parameters called ``signature
space'', the peak is expected to give a signal observable over
the SM prediction to a prescribed confidence level. Instead, the
{\it identification reach} on the model is the upper limit of the
range in $M_R$ where it can be identified as the source of the
peak,  once discovered, or, equivalently, the other competitor
models can be excluded for all values of their respective
parameters. Of course, for many models, identification should be
possible only in a subdomain of their signature space.
\par
The determination of the spin of an observed resonance clearly
represents an important selection among different classes of
nonstandard interactions. In the case of the inclusive diphoton
production (\ref{proc}), the tool to directly test the spin-2 of
the RS graviton resonance or, equivalently, exclude the hypothesis of
a spin-0 scalar particle exchange, would be provided by the
distinctive angular distributions in the angle $\theta$ between
the incident quark or gluon and the final photon in the diphoton
center-of-mass frame. This is similar to dilepton production, the
difference being that in this  case the hypotheses of both the
spin-0 and the spin-1 exchanges must simultaneously be
excluded.
\par
The spin-2 test of the lowest-lying RS graviton in lepton-pair
collider  events, through the direct comparison of the angular
distributions for the various spin hypotheses, was earlier
discussed in several papers, see, e.g.,
Refs.~\cite{Allanach:2000nr,Allanach:2002gn, Cousins:2005pq}, and experimental angular analysis were attempted at the Tevatron in
Ref.~\cite{Abulencia:2005nf}. A potential difficulty of  the
direct-fit angular analysis at the LHC is that generally, due to
the symmetry of the proton-proton initial configuration, the
determination on an event-by-event basis of the direction of the initial parton, hence of the sign of $\cos\theta$, is in principle not fully unambiguous, so that cuts in phase space must be applied in this regard.
\par
The spin-2 RS graviton analysis of LHC dilepton events proposed in
Ref.~\cite{Osland:2008sy}, makes use of a ``center-edge'' angular
asymmetry $A_{\rm CE}$ where the above mentioned ambiguity
should not be present  \cite{Dvergsnes:2004tc,Dvergsnes:2004tw}.
Essentially, in this observable the dilepton events are weighted
according to the $\cos\theta$ differential distributions, and the
asymmetry is defined between cross sections symmetrically
integrated over ``center'' and ``edge'' angular intervals.
Recently, asymmetries conceptually analogous to $A_{\rm CE}$, have
been applied to heavy quantum states spin identification in
Refs.~\cite{Diener:2009ee,DeSanctis:2011yc}, and a  comparison of
the performances of different methods for  heavy resonances
identification has been presented in Ref.\ \cite{Kelley:2010ap}. Angular analyses for
different spin-mediated Drell-Yan processes have been applied to a variety of NP models in
Ref.~\cite{Chiang:2011kq}.
\par
Here, we propose the application of $A_{CE}$ to the angular
analysis  of the diphoton production process (\ref{proc}) at the LHC. As remarked
previously, the selection of the spin-2 RS graviton amounts in
practice to exclude the hypothesis of a spin-0
particle exchange with same mass $M_R$ and
producing the same number of diphoton events. Ideally,
one advantage of the diphoton channel over dileptons can be
represented by the doubled statistics expected in the former case
\cite{Han:1998sg}. Also, the automatic exclusion of the spin-1
hypothesis~\cite{Landau:1948,Yang:1950}, should in any case
allow a simplification of the analysis from the phenomenological
point of view. Finally, the consideration of process (\ref{proc}),
in addition to  dilepton production, is needed for an exhaustive
test of model~\cite{Randall:1999ee}.
\par
For our analysis we have used the calculations of the required
differential cross sections to next-to-leading order (NLO) in QCD,
and this is essential at a hadron collider as the theoretical
uncertainties get reduced when higher order corrections are
included. Furthermore, as a result of new interactions in a NP
model, there will be additional subprocesses that contribute
at leading order (LO) itself (e.g., $gg \to\gamma \gamma$ in the RS model) and
hence the signal can receive enhanced contributions due to the
NLO corrections.
\par
Specifically, in Sec.\ II we review the definitions of the basic
cross sections involved in the asymmetry $A_{\rm CE}$; Sec.\ III
will be devoted to the relevant properties and  the characteristic
angular distributions for the RS graviton and for the competitor
scalar particle exchanges in process (\ref{proc}), for which we
will adopt the model recently proposed in Ref.\
\cite{Barbieri:2010nc}. In Sec.\ IV we discuss the NLO QCD effects
to the diphoton production rates and to the angular distributions,
for both kinds of spin exchange. Sec.\ V contains an outline of
the $A_{\rm CE}$-based angular analysis and the consequent
numerical results for RS identification, in the LHC center-of-mass
running configurations $\sqrt s=14$~TeV and $\sqrt s=7$~TeV.
Finally, Sec. VI contains some conclusive remarks.

\section{Cross sections and center-edge asymmetry}

The total cross section for a heavy resonance
discovery in the events (\ref{proc}) at a diphoton invariant mass $M=M_R$ can be expressed as
\begin{equation}
\sigma{(pp\to \gamma \gamma)}=\int_{-z_{\rm{cut}}}^{z_{\rm cut}}{\rm d} z
\int_{M_{R}-\Delta M/2}^{M_{R}+\Delta M/2}{\rm d} M  \frac{{\rm d}\sigma}
{{\rm d} M\, {\rm d} z},
\label{TotCr}
\end{equation}
where the rapidity of the individual photon $|\eta_\gamma|< 2.5$
and $z=\cos \theta$ is chosen such that $|z|<0.98$.
\par
Resonance spin-diagnosis uses the comparison between the
characteristic photon differential distributions for the two
hypotheses for the resonance bump, $R=G$ spin-2 Kaluza-Klein (KK) modes of the RS model
graviton and $R=S$ a massive scalar:
\begin{equation}
\frac{{\rm d}\sigma}{{\rm d} z} =
\int_{M_{R}-\Delta M/2}^{M_{R}+\Delta M/2}{\rm d} M
\frac{{\rm d}\sigma}{{\rm d} M\, {\rm d} z}.
\label{DiffCr}
\end{equation}
In Eqs.\ (\ref{TotCr}) and (\ref{DiffCr}), cuts on the phase space
accounting for detector acceptance are implicit, and $\Delta M$ is
an invariant mass bin around $M_R$, which should somehow reflect the
detector energy resolution and be sufficiently large as to include
the resonance width. In the calculations worked out in the sequel,
we use for this mass window the expression \cite{Feldman:2006wb}:
\begin{equation}
\Delta M=24\left(0.625 M+M^2+0.0056\right)^{1/2}\,
{\rm GeV}.
\label{deltam}
\end{equation}
Actually, Eq.(\ref{deltam}) was derived in
connection to the ATLAS and CMS experiments on
dilepton production, but we assume it also for
the calculations of diphoton production of
interest here. Obviously, for a resonance
sufficiently narrow, the integral over
$M$ should be practically insensitive
to the size of $\Delta M$, whereas it should
be essentially proportional to $\Delta M$ for
a flat background such as the SM.
Besides $\vert\eta_\gamma\vert <2.5$ mentioned above, the assumed typical
cuts on harder (softer) photons are
$p^\gamma_\bot >40 (25)$~GeV,
and the statistics will be estimated by taking
a photon reconstruction efficiency
$\epsilon_\gamma=0.80$.
\par
Moreover, to evaluate Eqs.\ (\ref{TotCr}) and
(\ref{DiffCr}), the partonic cross sections will
be convoluted with the CTEQ6L and CTEQ6M parton
distributions sets for LO and NLO cross
sections, respectively, with
$\Lambda_{QCD}=0.226$ GeV \cite{Pumplin:2002vw}.
In particular, for resonance discovery,
process (\ref{proc}) must be observed with a
number of events well-above the background from SM
processes. Specifically, denoting by $N_S$ and
$N_{\rm SM}$ the numbers of signal and SM
events in the $\gamma\gamma$ invariant mass
window, the statistical significance of a
5-$\sigma$ signal would be ensured by
the criterion that $N_S$ should be larger than
max($5\sqrt{N_{\rm SM}}$,$\,10$).
\par
The $z$-evenly-integrated center-edge angular
asymmetry $A_{\rm CE}$ is defined as:
\begin{equation}
\label{ace}
A_{\rm{CE}}=\frac{\sigma_{\rm{CE}}}{\sigma}\quad{\rm
with} \quad \sigma_{\rm{CE}} \equiv \left[\int_{-z^*}^{z^*} -
\left(\int_{-z_{\rm cut}}^{-z^*} +\int_{z^*}^{z_{\rm
cut}}\right)\right] \frac{{\rm d} \sigma}{{\rm d} z}\, {\rm d} z.
\end{equation}
In Eq.\ (\ref{ace}), $0<z^*<z_{\rm cut}$ defines the separation
between the ``center'' ($|z|<z^*$) and the ``edge''
($z^*<|z|<z_{\rm cut}$) angular regions and is {\it a priori}
arbitrary to some extent. In previous applications, see for example
Refs.\ \cite{Osland:2008sy,Dvergsnes:2004tc,Dvergsnes:2004tw,
Diener:2009ee,DeSanctis:2011yc}, the ``optimal'' numerical value
turned out to be $z^*\simeq 0.5$, and we shall keep this value of
$z^*$ here as well. One can notice that by definition $A_{\rm CE}$
is symmetric under $z\leftrightarrow -z$, hence it is insensitive
to the sign of $z$. Moreover, as being a ratio of integrated cross
sections, an advantage of $A_{\rm CE}$ is that it should be less
sensitive to theoretical systematic uncertainties, such as the
uncertainties from different sets of parton distributions and from
the particular choice of factorization and renormalization scales.


\section{Graviton resonance and scalar exchanges}

We here sketch the models we are interested in,
together with their features relevant to the
resonance spin and the distinctive angular
distributions for process (\ref{proc}).

\subsection{RS model of gravity with one compactified extra dimension}

This model, originally proposed as a solution to the gauge
hierarchy problem $M_{\rm EW}\ll M_{\rm Pl}$, consists of two
3-branes, and one compactified warped extra spatial dimension $y$
with exponential warp factor $\exp{(-k\pi\vert y\vert)}$
\cite{Randall:1999ee}.  Here, $k>0$ is the 5D curvature, assumed
to be of the order of $M_{\rm Pl}$.  The two branes are placed at
orbifold fixed points, $y=0$ with positive tension called the
Planck brane and the second brane at $y=R_c$ with negative tension
called the TeV brane. The basic, simplifying, hypothesis is that
the SM fields are localized on the TeV brane, whereas gravity
originates on the Planck brane but is allowed to propagate
everywhere in the 5D space. The consequence of this setup is the
existence of KK modes of the graviton that can be exchanged in the
interactions among SM particles in TeV brane.  Owing to the
exponentially suppressing warp factor, mass scales, in passing from
the Planck brane to the TeV brane, can get the size of the TeV
order.  Moreover, a specific mass spectrum of such KK resonances
is predicted in terms of an effective mass scale defined as
$\Lambda_\pi={\overline M}_{\rm Pl}\exp{(-k\pi R_c)}$, that for $k R_c\simeq 12$ happens to be of the TeV order
(here, ${\overline M}_{\rm Pl}=1/\sqrt{8\pi G_{\rm N}}$ with
$G_{\rm N}$ the Newton constant). These resonances, represented by
spin-2 fields ${h^{(n)}_{\mu\nu}}$, can in process (\ref{proc})
show up as (narrow) peaks in $M_{\gamma\gamma}\equiv M$, through
the interaction
\begin{equation}
{\cal L} = -\frac{1}{\Lambda_{\pi}} T^{\mu\nu} \sum_{n=1}^{\infty}
{h^{(n)}_{\mu\nu}}.
\end{equation}
Here: $T^{\mu\nu}$ is the energy-momentum tensor of the SM; the
characteristic mass spectrum is $M_n=x_nk\exp{(-k\pi R_c)}$ (with
$x_n$ the roots of the Bessel function $J_1(x_n)=0$); and the
resonance widths are $\Gamma_n = \rho M_n x^2_n(k/{\overline M}_{\rm Pl})^2$,
with $\rho$ a calculable constant depending on the number of open decay
channels, of the order of $0.1$.
\par
The model can therefore conveniently be parametrized in terms of
$M_G\equiv M_1$, the mass of the lowest graviton excitation, and
of the ``universal'' dimensionless graviton coupling
$c=k/{\overline M}_{\rm Pl}$. Theoretically, the expected ``natural''
ranges for these parameters, avoiding additional mass hierarchies,
are: $0.01< c < 0.1$ and $\Lambda_\pi< 10$ TeV~\cite{Davoudiasl:2000wi}.
The 95\% CL experimental lower bounds on $M_G$ from previous analysis vary, essentially, from $0.6$ to $1.0$ TeV as $c$ ranges from $0.01$ to $0.1$ \cite{Aaltonen:2010cf,Abazov:2010xh,Chatrchyan:2011wq}.
Quite recently, preliminary results from RS graviton searches in dilepton inclusive production at the 7 TeV LHC with luminosity 1.2 ${\rm fb}^{-1}$, indicate
95\% CL lower limits on $M_G$ of 0.7 TeV for $c=0.01$
up to 1.6-1.7 TeV for $c=0.1$~\cite{atlas2,cms2}.
\par
\begin{figure}[!htb]
\begin{center}\vspace*{0.0cm}
\includegraphics[width=10.0cm,angle=0]{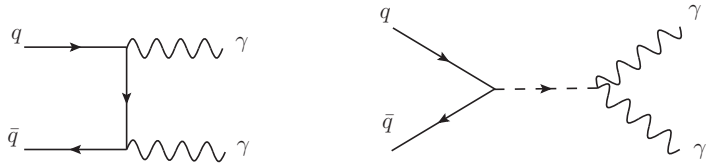}
\caption{Feynman diagrams contributing to the subprocess
$q \bar{q} \rightarrow \gamma \gamma$, where the dashed
line represents either a RS graviton KK mode $G$ or a scalar
exchange $S$. The crossed diagrams are not displayed.}
\label{qq}
\end{center}
\end{figure}
\begin{figure}[!htb]
\begin{center}\vspace*{0.0cm}
\includegraphics[width=10.0cm,angle=0]{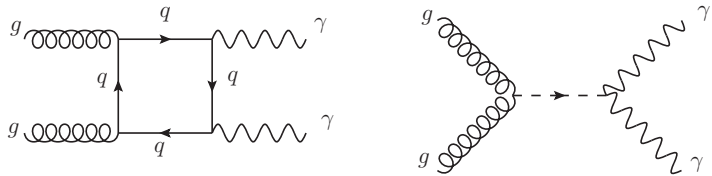}
\vspace*{-0.0cm} \caption{Feynman diagrams contributing to the subprocess
$gg \rightarrow \gamma \gamma$, including the exchange of either a RS
graviton KK mode $G$ or a scalar $S$ which is denoted by the dashed line.
The crossed diagrams are not displayed.}
\label{gg}
\end{center}
\end{figure}

In hadronic collisions, in QCD at LO, photon pairs can be produced {\it via}
the quark--antiquark annihilation $q+\bar{q}\to \gamma +\gamma$, and the
gluon--gluon fusion $g+g \to \gamma +\gamma$. The relevant diagrams at this
order, for the SM and the RS graviton exchange, are represented in Figs.\
\ref{qq} and \ref{gg}. Actually, the SM box diagram in Fig.\ \ref{gg} is of
higher order in $\alpha_s$ and, as discussed in the next section, for the
values of the $\gamma\gamma$ invariant mass $M$ in the TeV range of interest
here, its contribution turns out to be negligible, as earlier noticed also in
Ref.\ \cite{Eboli:1999aq}.

\par
Using Feynman rules for graviton exchange \cite{Han:1998sg,Giudice:1998ck},
the $z$-even angular dependencies needed in (\ref{ace}) can be written
as~\cite{Cheung:1999wt,Sridhar:2001sf,Giudice:2004mg}:
\begin{eqnarray}
\frac{\dd\sigma (q\bar{q}\to\gamma\gamma)}{\dd z} &&=
\frac{1}{192\,\pi\hat{s}} \Bigl[ 64\,\alpha^2 \pi^2\, Q^4_q\,
\frac{1+z^2}{1-z^2}  
+
\frac{\hat{s}^4}{16} \, |C(\hat{s})|^2\, (1-z^4)
\Bigr.
\nonumber \\
&&
- \Bigl.
\ {4 \pi\alpha\, Q^2_q ~\hat{s}^2~ {\rm Re}
[C(\hat{s})]}\left(1+z^2\right)
\Bigr]\ ,
\label{dsigmaqq}
\end{eqnarray}
where the subprocess Mandelstam variable is the
diphoton invariant mass, $\sqrt{\hat{s}}\equiv M$,
and $\alpha$ is the electromagnetic coupling
constant with $Q_q$ the quark electric charge.
Moreover, in Eq.~(\ref{dsigmaqq}), $C(\hat{s})$
represents the sum of KK graviton propagators 
with masses $M=M_n$ and widths $\Gamma_n$:
\begin{equation}
C(\hat{s})= \frac{1}{\Lambda^{2}_{\pi}}\sum_n\frac{1}
{\hat{s}-M^{2}_{n}+iM_{n}\Gamma_{n}}. \label{gravipropagator}
\end{equation}
In practice, from the phenomenology, just the
lowest graviton mass $M_G\equiv M_1$ and,
perhaps, the next one $M_2$ at most,
can be expected to fall within the
discovery reach of LHC.
\par
The cross section for the $gg\to\gamma\gamma$ subprocess {\it via} the RS
graviton excitation exchange is
\begin{equation}
\frac{\dd\sigma (gg \to\gamma \gamma)}{\dd z} =
\frac{\hat{s}^3}{8192\, \pi} \,
|C(\hat{s})|^2\, (1+ 6 z^2 + z^4).
\label{dsigmagg}
\end{equation}
Notice that a factor $1/2$ is embodied in Eqs.\ (\ref{dsigmaqq}) and (\ref{dsigmagg})
to account for the identical final state photons.

\subsection{The model for scalar exchange}

In principle, in order to discriminate the graviton  spin-2
angular distribution from the spin-0 hypothesis, we might limit
ourselves to make a comparison of Eqs.~(\ref{dsigmaqq}) and
(\ref{dsigmagg}) with the results of a generic flat (in $z$)
distribution numerically tuned to the same number of events around
$M_G$. However, for the graviton exchange we shall use a
description supplemented by the cross sections calculated to
NLO in QCD and, to consistently fully exploit
the QCD dynamics also for the spin-0 scenario, we need an explicit
model for a scalar particle exchange with definite couplings to
the SM fields, in particular to photon pairs.
\par
We consider the simple model of a scalar particle $S$,  singlet
under the SM gauge group and with mass $M\equiv M_S$ of the TeV
order, proposed in Ref.~\cite{Barbieri:2010nc}. The trilinear
couplings of $S$ with gluons, electroweak gauge bosons and
fermions, are in this model:
\begin{equation}
{\cal L}=
c_3\frac{g_s^2}{\Lambda}G^a_{\mu\nu}G^{a\ \mu\nu}S+
c_2\frac{g^2}{\Lambda}W^i_{\mu\nu}W^{i\ \mu\nu}S+
c_1\frac{g^{\prime 2}}{\Lambda}B_{\mu\nu}B^{\mu\nu}S+
\sum_f c_f\frac{m_f}{\Lambda}{\bar f}f S.
\label{scalar}
\end{equation}
In Eq.~(\ref{scalar}), $\Lambda$ is a high mass
scale, of the TeV order of magnitude, and $c$'s
are dimensionless coefficients that are assumed
to be of order unity, reminiscent of a strong novel
interaction. In our subsequent analysis we
take, following Ref.~\cite{Barbieri:2010nc},
$\Lambda=3$ TeV.
As for the $c$'s, we
shall leave their numerical values free to vary
in a range of the order of (or less than) unity, but
constrained so that the scalar particle width
$\Gamma_S$ could be comparable to (or included in)
the diphoton mass window
$\Delta M$ of Eq.~(\ref{deltam}).
\par
The leading order diphoton production process is in this  model dominated by the $s$-channel exchange 
$gg\to S\to\gamma\gamma$. As it is of
order $\alpha_s^2$ at LO, it will be sensitive to the choice of this
coupling constant.
The Feynman diagrams in the scalar model will be similar to those in the
RS model, as shown in Figs.\ \ref{qq} and Fig.\ \ref{gg} with the KK mode
replaced by the scalar. The corresponding differential cross section reads, including a factor $1/2$ for identical final particles:
\begin{equation}
\frac{\dd\sigma (gg\to\gamma \gamma)}{\dd z} =
\frac{1}{2}
\frac{1}{16\pi}\ \left(\frac{c_3g_s^2}{\Lambda}\right)^2\
\left(\frac{(c_1+c_2)e^2}{\Lambda}\right)^2\
{\hat s}^3\ \vert D({\hat s})\vert^2.
\label{dsigmaggs}
\end{equation}
In Eq.~(\ref{dsigmaggs}), $D({\hat s})$ is the
scalar propagator,
\begin{equation}
D({\hat s})=
\frac{1}{{\hat s}-M_S^2 + i M_S\Gamma_S},
\label{propscal}
\end{equation}
and the expression of the total width $\Gamma_S$ in terms of the
$c$'s and $\Lambda$ introduced in Eq.~(\ref{scalar}) can be
obtained from   \cite{Barbieri:2010nc},  by summing the partial
widths reported there into $gg$ (dominant, as is the cross
section, proportional to $c_3$), $\gamma\gamma$,  $W^+W^-$, $ZZ$, and ${\bar f}f$.
\par
\begin{figure}[tbh!] 
\vspace*{0.5cm}
\centerline{ 
\includegraphics[width=8.0cm,angle=0]{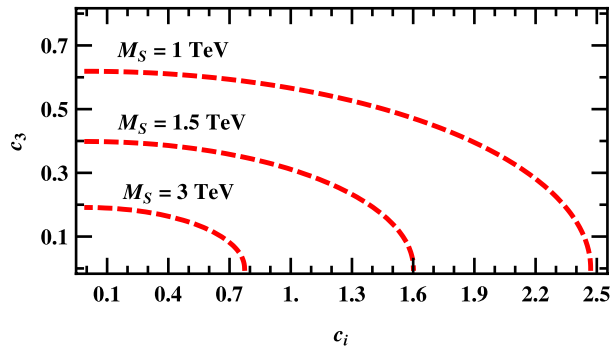}}
\caption{\label{fig-narrow-scalar} The dashed lines in the
parameter plane $(c_i,c_3)$ with equal $c_i$ ($i\neq 3$),
correspond to the equation $\Gamma_{S} = \Delta M$ for $M_S=1$
TeV, 1.5 TeV and 3 TeV, $\Lambda = 3$~TeV. The allowed area is at
the left of these lines.}
\end{figure}

In Fig.~\ref{fig-narrow-scalar} 
we represent, as an example, the
domains in $c_3$ {\em vs} all other parameters 
$c_i$ ($i\neq 3$) assumed equal to each other, 
allowed by the constraint $\Gamma_S\leq \Delta M$ 
for different values of $M_S$.
\par
As the coupling of the scalar particle to quarks is proportional to the quark mass much
smaller than $\Lambda$, one can neglect at LO the subprocess ${\bar q}q\to S\to\gamma
\gamma$, and the cross section for $pp\to S\to\gamma\gamma$ will be expressed simply by
the gluon-initiated process plus the SM contribution appearing in Eq.\ (\ref{dsigmaqq}).
Thus, at this LO, there is in the considered spin-0 model no interference between the
SM quark-initiated contribution and the scalar-exchange amplitude.  However, quark
contributions cannot be neglected at the NLO QCD order and, in particular,
a (small) interference of the $gg$ SM box diagram with the $gg$ scalar exchange will
occur, as it will be specified in the next section.
\par
Preliminary to the numerical analysis of $A_{\rm CE}$, 
in the next section we briefly describe the
estimate of the next-to-leading order QCD effects.

\section{Next-to-leading order QCD effects}

The diphotons produced in a hadronic collision could originate from the hard partonic
interaction (direct photon) or at least one photon could be produced in the hadronisation
of a parton (fragmentation photon). At higher orders in QCD, there would be final state
collinear singularities of QED origin as a result of the emission of a photon from a quark.
These singularities can be factored out and absorbed into the fragmentation functions. The
fragmentation functions are additional non perturbative inputs that are not well-understood.
An alternate approach to isolate direct photons is the smooth cone isolation criterion
\cite{Frixione:1998jh}, which ensures that the fragmentation contributions are suppressed
without affecting the cancellation of the conventional QCD singularities.
\par
We start from the extra dimension scenario, where the NLO corrections
in QCD were considered for the phenomenologically interesting process like
the dilepton \cite{Mathews:2004xp,Mathews:2005bw,Mathews:2005zs} and
diphoton production \cite{Kumar:2009nn,Kumar:2008pk} at hadron colliders.
The essential Feynman rules for KK modes coupling to ghosts and KK mode
propagator in $(4+\epsilon)$-dimensions needed for the NLO computation
have been introduced in \cite{Mathews:2004xp}.
For the diphoton computation a semi-analytical two-cutoff phase space
slicing method \cite{Harris:2001sx} to deal with various singularities of
infrared (IR) and collinear origin that appear at NLO in QCD was used.
After cancellation and mass factorisation of these singularities,
the remaining finite part is numerically integrated over the phase
space by using Monte-Carlo techniques.
\par
As stated in previous sections, diphotons can in the SM be
produced to LO in QCD, in the quark anti-quark annihilation
subprocess $q \bar q \to \gamma \gamma$. Photon pairs produced {\it via} 
the gluon fusion subprocess through a quark box diagram $g g \to
\gamma \gamma$, though of the order $\alpha_s^2$, have cross
sections comparable to those of the $q \bar q \to \gamma \gamma$
subprocess for the low diphoton invariant mass.  In the light
Higgs boson searches, this subprocess plays an important role, due
to the large gluon flux at small fractional momentum $x$, and is
formally treated as a LO contribution although it is of ${\cal O}
(\alpha_s^2)$, hence is in reality a next-to-next-to leading order
contribution.  However, it falls off rapidly with increasing
diphoton invariant mass and in the mass range of interest for the
TeV scale gravity models it need not be included at LO. It has
been demonstrated in \cite{Kumar:2009nn,Kumar:2008pk} that the
contribution of this subprocess in the SM is few orders of
magnitude smaller than that of the $q \bar q$ sub process
for diphoton invariant mass $M > 500$ GeV.\footnote{See also  the discussion of the SM NLO predictions for process 
(\ref{proc}) in Ref.~\cite{Campbell:2011bn}.} 

\begin{figure}[!htb]
\begin{center}\vspace*{0.0cm}
\includegraphics[width=13.0cm,angle=0]{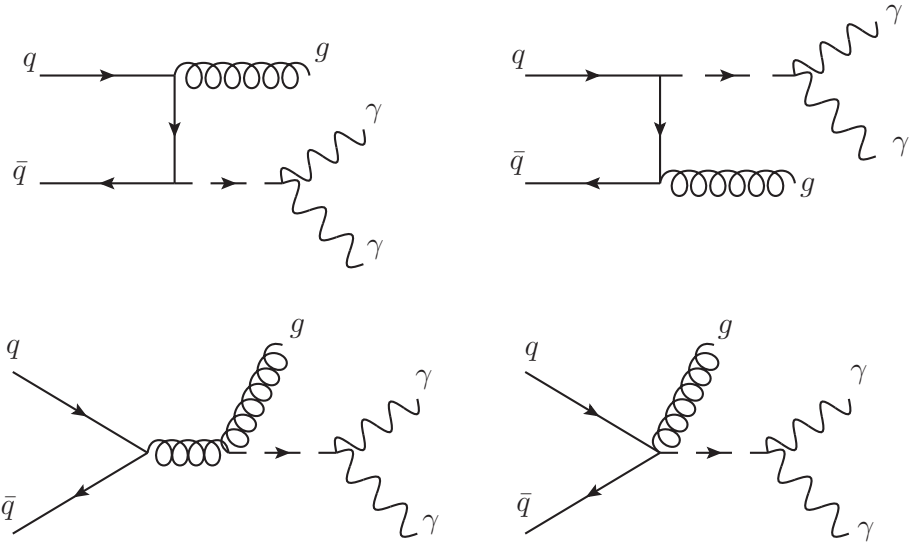}
\vspace*{0.0cm} \caption{Feynman diagrams contributing to the subprocess
$q \bar q \rightarrow \gamma \gamma g$.  The dashed line could be either $R=G$
or $S$ depending on the model. For the scalar exchange $S$ the last diagram
involving a four point vertex does not exist.}
\label{qqreal}
\end{center}
\end{figure}
\begin{figure}[!htb]
\begin{center}\vspace*{0.0cm}
\includegraphics[width=13.0cm,angle=0]{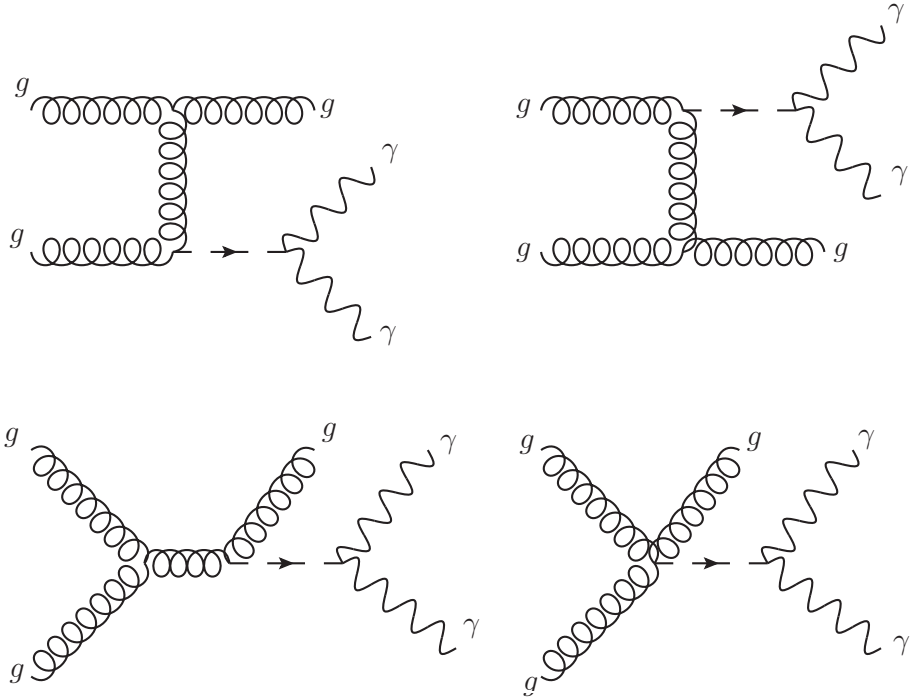}
\vspace*{0.0cm} \caption{Feynman diagrams contributing to the
subprocess  $g g \rightarrow \gamma \gamma g$, where the dashed
line corresponds to a spin-2 RS KK mode $G$ or a scalar exchange
$S$.  In the case of a scalar exchange the last diagram involving
a four point vertex does not exist.}
\label{ggreal}
\end{center}
\end{figure}

\par
\begin{figure}[!htb]
\begin{center}\vspace*{0.0cm}
\includegraphics[width=13.0cm,angle=0]{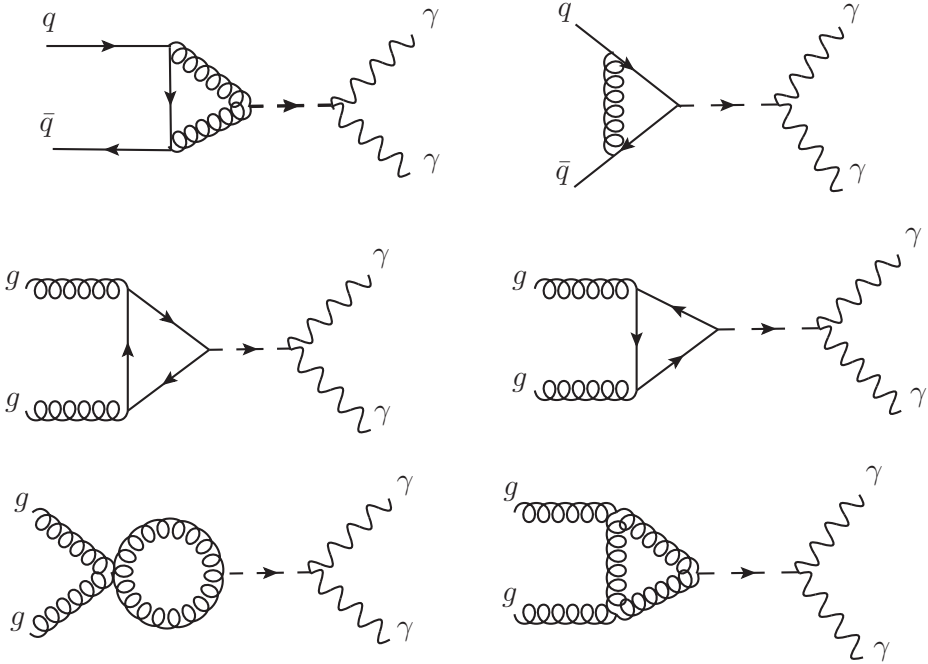}
\vspace*{0.0cm} \caption{Feynman diagrams contributing to
one loop subprocess to the $q \bar q$ and $gg$ channel in
the NP models.}
\label{virtual}
\end{center}
\end{figure}

The NLO QCD corrections to the diphoton production process would
involve real emission diagrams corresponding to the following
$2 \to 3$ subprocesses:
(a) $q \bar q \to \gamma \gamma g$ (Fig.\ \ref{qqreal}),
(b) $g g \to \gamma \gamma g$ (Fig.\ \ref{ggreal})
and (c) $g q (\bar q) \to \gamma \gamma q (\bar q)$
(the $g q$ diagrams can be obtained from the $q \bar q$
diagram by appropriate inter change of initial and final
state particles).  In the virtual part, ${\cal O} (\alpha_s)$
corrections at one loop arise as a result of the interference
between the one loop graphs at ${\cal O} (\alpha_s)$ of
$(SM + NP)$ and the Born graphs at ${\cal O} (\alpha_s^0)$ of
$(SM + NP)$. Some of the typical NP one loop Feynman diagrams to
${\cal O} (\alpha_s)$ are shown in Fig.\ \ref{virtual}.  The
$q \bar q$ channel gets such contributions
from both the SM and the graviton exchange, while in the $gg$
channel the SM contribution already begins at ${\cal O} (\alpha_s)$.
The SM $g g \to \gamma \gamma$ subprocess amplitude can interfere
with the gluon initiated LO gravity exchange subprocess, giving a
${\cal O} (\alpha_s)$ contribution Fig.\ \ref{gg}.  In the $qg$
channel there is no virtual contribution to this order in either
the SM or the NP models of interest here.
\par
Both the virtual and real corrections have been evaluated with 5 quark flavors and in
the limit of vanishing quark masses. The $n$-point tensor integrals appearing from
integration over loop momenta were simplified using the Passarino-Veltman reduction
\cite{Passarino:1978jh}, the computation was done using dimensional regularization in
$(4+\epsilon)$-dimensions, in the $\overline {MS}$ scheme.
\par
The virtual contribution here does not
contain UV singularities, this can be attributed
to the facts that {\it i)} the electromagnetic
current is conserved and does not receive any
QCD corrections, and {\it ii)} that the
graviton couples to the energy-momentum
tensor of the SM fields which
is a conserved quantity and therefore is
not renormalized. The poles in $\epsilon$
arise from loop integrals and correspond to
the soft and collinear divergences,
configurations where a virtual gluon momentum
goes to zero give soft singularities while
the collinear singularities arise
when two massless partons become collinear to
each other.
\par
The 3-body phase space of the real emission diagrams has regions
which are soft and collinear divergent.  The phase space can
accordingly be separated into soft $(s)$ and hard regions and,
furthermore, the hard region can be separated into hard collinear
$(HC)$ and hard non-collinear $(\overline {HC})$ parts as follows:
\begin{eqnarray}
d \sigma^{real}=d \sigma^{real}_s(\delta_s)
+ d \sigma^{real}_{HC} (\delta_s,\delta_c)
+ d \sigma^{real}_{\overline {HC}} (\delta_s,\delta_c) ~.
\label{phs1}
\end{eqnarray}
The small cut-off parameters $\delta_s$ and $\delta_c$ set arbitrary
boundaries for the soft (gluon energy $0\le E_g\le \delta_s \sqrt{\hat s}/2$)
and collinear ($0\le \hat t_{ij}\le\delta_c \hat s$) regions, respectively.
Here, for the $2 \to 3$ process with momenta $p_1+p_2 = p_3 +p_4+p_5$, the
Mandelstam variables are defined as $\hat s_{ij}=(p_i+p_j)^2$, $\hat t_{ij}=
(p_i -p_j)^2$, with $\hat s = \hat s_{12}$.  In these mutually exclusive
soft and hard collinear regions, the 3-body cross section can be factored
into the Born, $2\to 2$, cross section and the remaining phase space integral
can easily be evaluated in $(4+\epsilon)$-dimensions to get the expansion of
the soft and collinear singularities in powers of $\epsilon$.  All positive
powers of the small cut-off parameters $\delta_s$ and $\delta_c$ are set to
zero, only logarithms of the cut-off parameters are retained.  Adding the
virtual and real contributions, all double and single poles of IR origin are
automatically cancelled between the virtual
and the first two terms of Eq.~(\ref{phs1}).
The remaining collinear poles are absorbed
into the parton distribution functions.
\par
The hard non-collinear part in Eq.\ (\ref{phs1}) corresponds to the 3-body
phase space, which by construction is finite, and can be evaluated in
4-dimensions ($\epsilon=0$). The sum of real and virtual contributions is
now free of singularities and can hence be evaluated numerically using a
Monte-Carlo method.  It can be further seen that the explicit logarithmic
dependence on $\delta_s$ and $\delta_c$ in the 2-body phase space
($d \sigma^{real}_s(\delta_s) + d \sigma^{real}_{HC} (\delta_s,\delta_c)$)
is cancelled by the implicit dependence of the 3-body hard non-collinear
part $d \sigma^{real}_{\overline {HC}} (\delta_s,\delta_c)$ on these
parameters, after the numerical integration is carried out. The sum of the
2-body and 3-body parts in Eq.\ (\ref{phs1}) would be independent of the
slicing parameters $\delta_s$
and $\delta_c$, which is explicitly verified before the code is used for
the analysis.  Now these codes \cite{Kumar:2009nn,Kumar:2008pk} can be used
to study the full quantitative implication of the NLO QCD corrections to
the various distributions of interest in the extra dimension searches.

\par
Turning to the scalar-exchange model, the matrix elements corresponding
to the interference between the SM box diagram and the LO $gg$ initiated
tree level diagram is given by
\begin{eqnarray}
I^{gg}_{SM*S} (z) & = & \frac{g_s^2}{16 \pi^2} Q_q^2 Re\left[D(\hat{s})\right]
{1 \over [N^2-1]} ~\left({c_3~g_s^2} \over \Lambda \right) ~
\left( {(c_1+c_2)~e^2 \over \Lambda }\right)~
\nonumber \\
{} && \times \Bigg\{
\left( {\hat{t}-\hat{u} \over 2}\right)
\left[\hbox{ln}\left({-\hat{t} \over \hat{s}}\right) -
\hbox{ln}\left({-\hat{u} \over \hat{s}}\right) \right]
\nonumber \\
{} && -\left( {{\hat{u}^2+\hat{t}^2} \over 2s} \right)
\left[ \hbox{Li}_2\left({-\hat{u} \over \hat{t}}\right) +
\hbox{Li}_2\left({-\hat{t} \over \hat{u}}\right) -2~\zeta(2) \right]
\Bigg\},
\end{eqnarray}
where $\hat{s}$, $\hat{t}= -{\hat{s} \over 2 } (1-z)$, $\hat{u} = -{\hat{s}\over 2}(1+z)$
are the usual Mandelstam variables, $N$ is the colour degrees of freedom and
$\zeta(2)=\pi^2/6$ is the Riemann zeta function.
\par
In the scalar model, the UV divergences in the virtual corrections
to the $gg$ initiated diphoton production process are removed by
renormalization.  The remaining finite contribution is given by
\begin{eqnarray}
\frac{d\sigma_v} {dz} =  \frac{g_s^2}{16 \pi^2} ~8N~\zeta(2)~ \frac{d\sigma}
{dz}
\end{eqnarray}
where $d\sigma/dz$ is the Born contribution given in Eq.\ (\ref{dsigmaggs}). The
remaining soft and collinear finite contributions coming from the real corrections
to the LO scalar model diagram are proportional to the Born cross section. As they
originate from pure QCD, they are independent of the hard scattering process, hence
they will be the same as those for the $s$-channel diphoton production process in
the RS model and are given in Ref.\ \cite{Kumar:2009nn}.
\par
At the NLO in QCD, the following three subprocesses contribute to the scalar model cross
section: {\it i)} ${\bar q}q\to gS$ {\it ii)} $qg\to qS$ and {\it iii)} $gg\to gS$, all
followed by $S \to \gamma\gamma$.  The Feynman diagrams for the $\bar q q$ and $gg$ channels are
given Figs.\ \ref{qqreal} and \ref{ggreal}, respectively, wherein the dashed
line now represents the scalar $S$ and the four point diagrams will be absent
for the scalar case.  The couplings of $S$ to the light quarks are proportional
to the masses and hence are negligible so that the main contribution to the
$\bar q q$ channel is from the $S$ coupling to the gluons.  Again the $q g$
channel is related to the  $\bar q q$ channel by suitable change of initial
and final state.
\par
\begin{figure}[ptbh!] 
\vspace*{0.5cm} \centerline{ \hspace*{0.5cm}
\includegraphics[width=9.0cm]{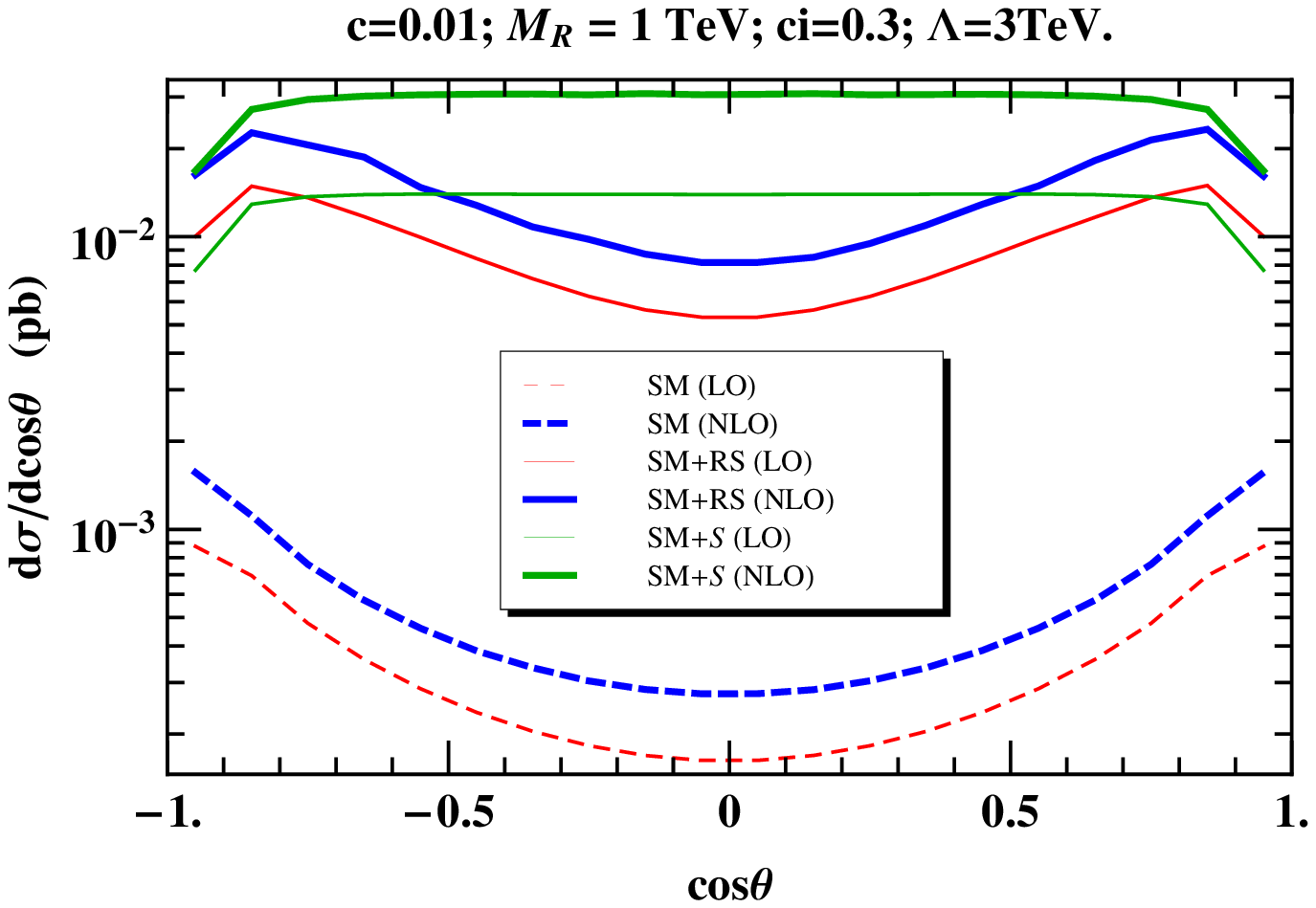}
\hspace{-0.9cm}%
\includegraphics[width=9.0cm]{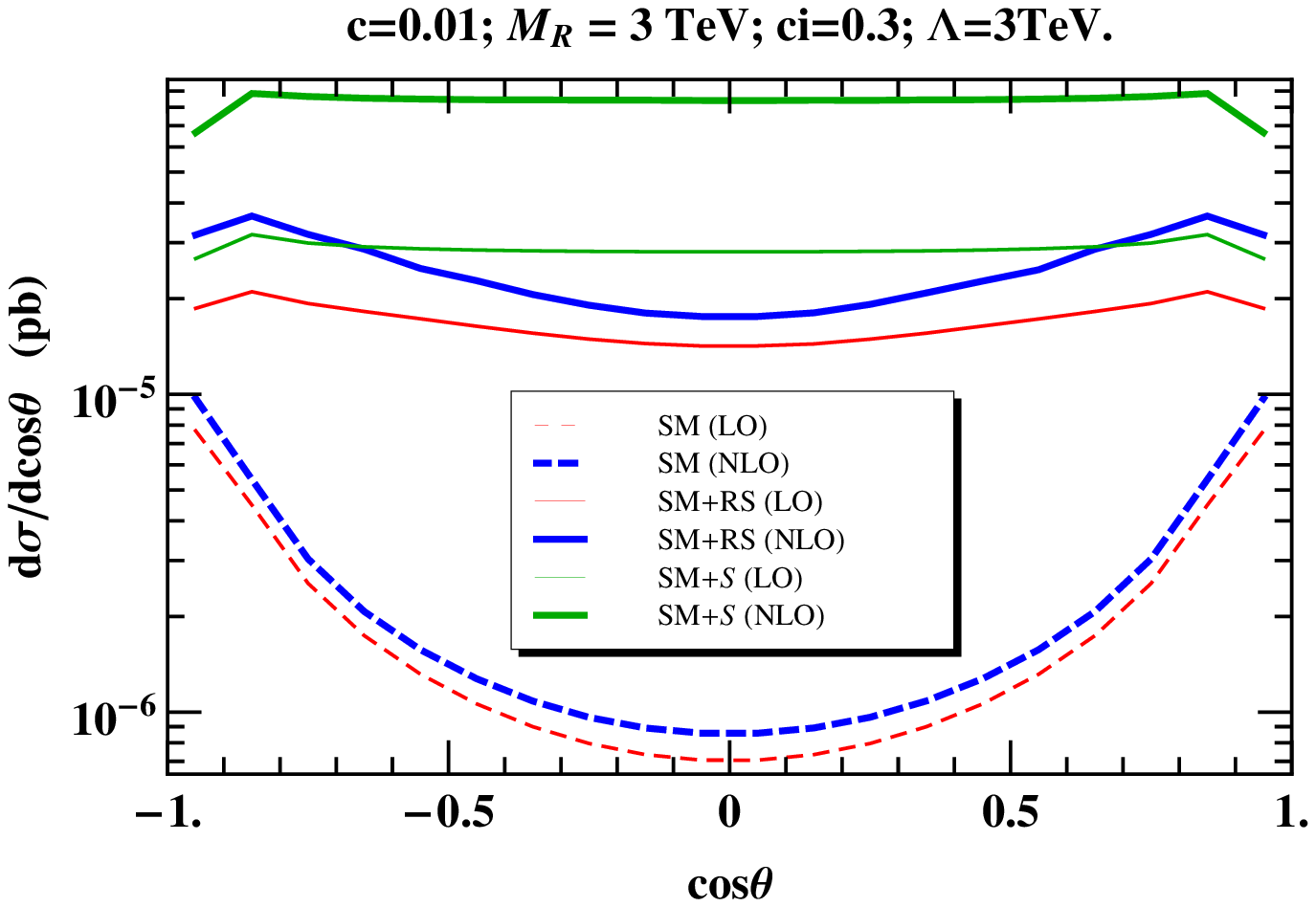}}
\centerline{ \hspace*{0.5cm}
\includegraphics[width=9.0cm]{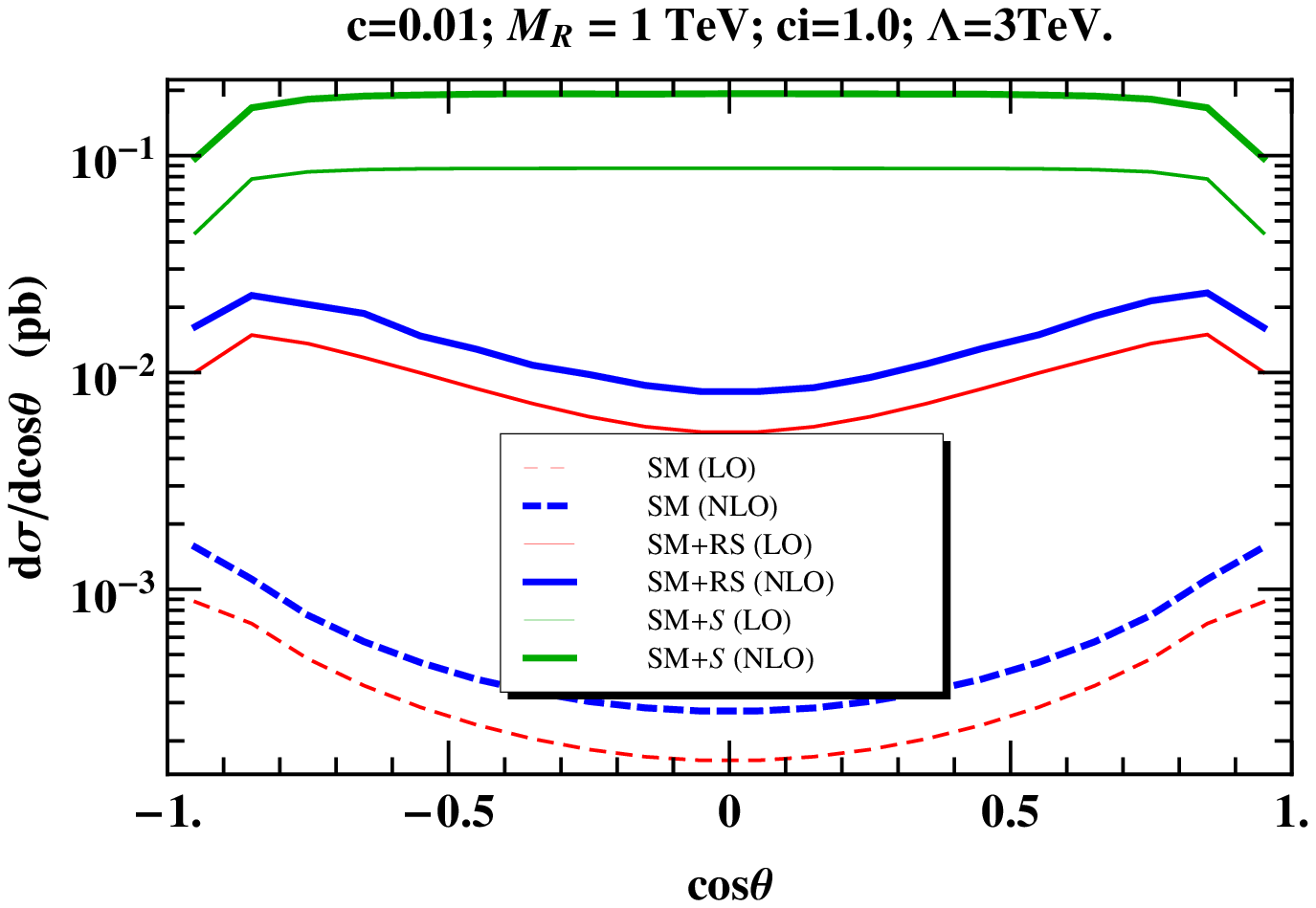}
\hspace{-0.9cm}%
\includegraphics[width=9.0cm]{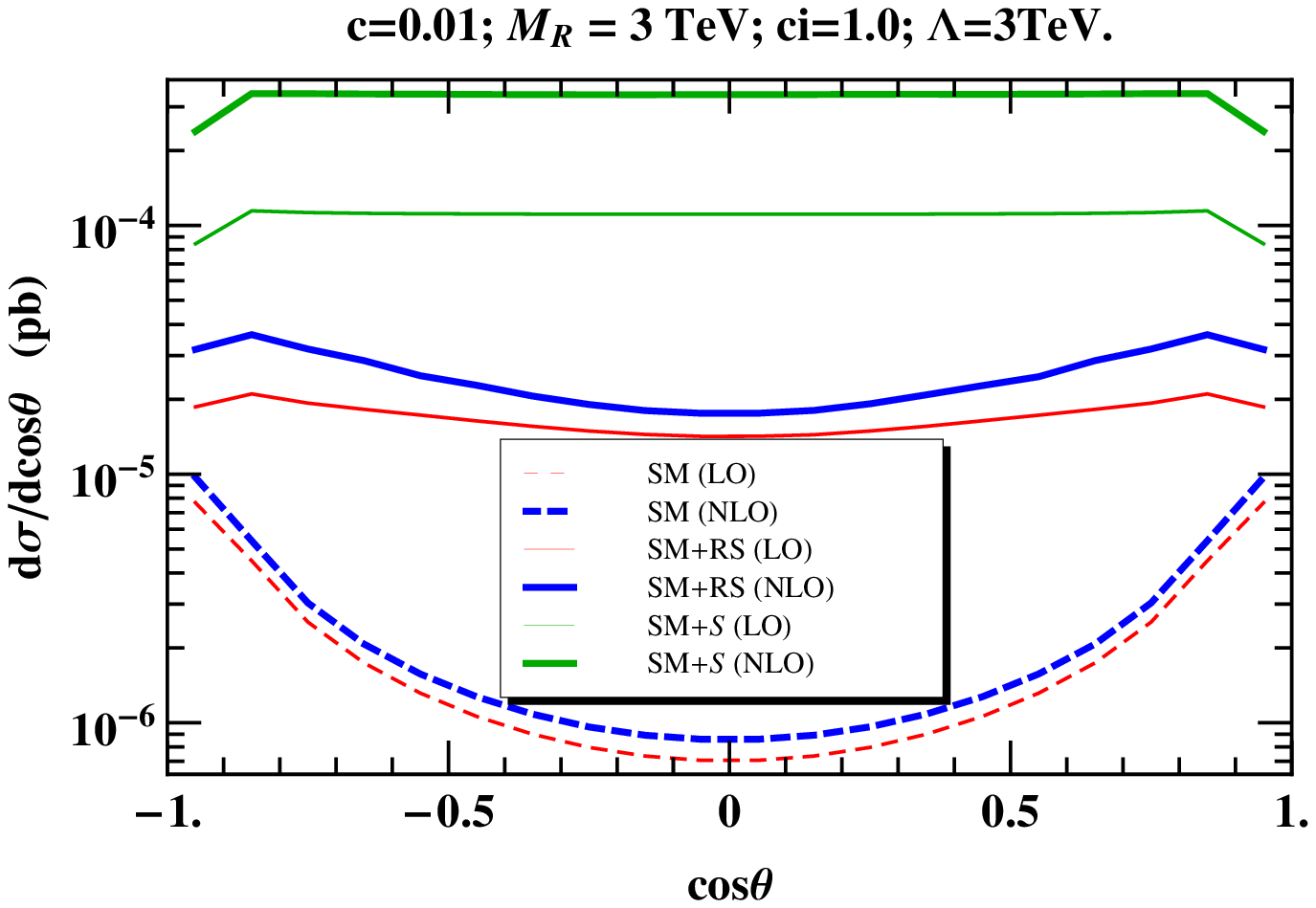}
}
\caption{\label{fig-ang-distr} Angular distributions of photons
for process $p+p\to \gamma\gamma+X$ at LO and NLO QCD in SM, RS
model with graviton coupling constant $c=0.01$, and scalar
resonance model with $c_3=0.3$, $c_i=0.3$, $i\neq 3$ (left and right top
panels) and $c_3=0.3$, $c_i=1.0$ (left and right bottom panels).
$M_R=1$ TeV and 3 TeV at the 14~TeV LHC. }
\end{figure}

Fig.\ \ref{fig-ang-distr} shows, as an example, for the
diphoton resonance mass
values $M_R=1$ and 3 TeV, the angular distributions at the 14 TeV LHC, at LO and
NLO, in the cases of the SM, the RS graviton exchange with coupling constant $c=0.01$,
and the scalar particle exchange with couplings ($c_3=c_i=0.3$) and ($c_3=0.3, c_i=1.0$), $i\neq 3$.

\begin{figure}[ptbh!] 
%
\centerline{ \hspace*{-0.4cm}
\includegraphics[width=10.3cm]{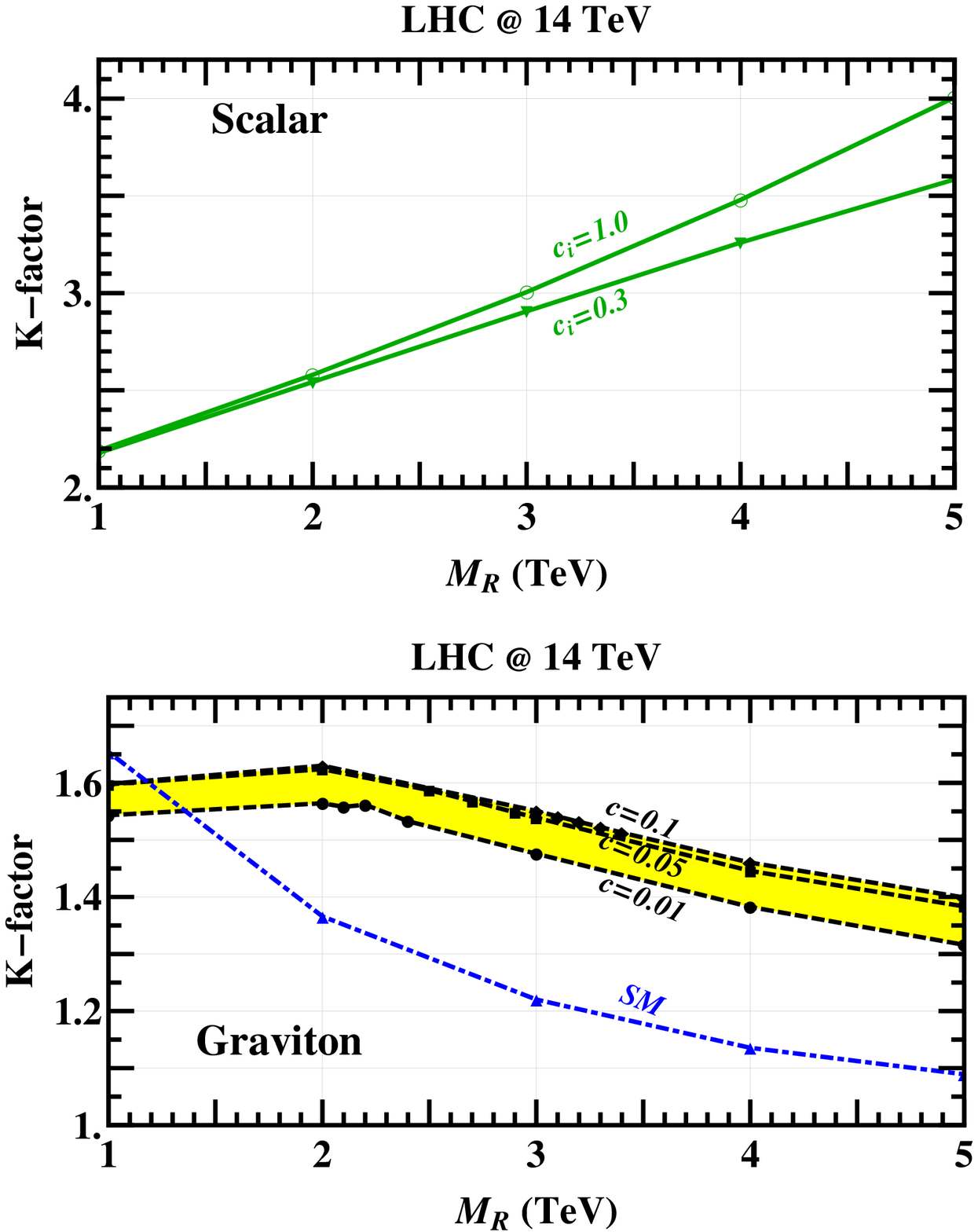}
\hspace*{-2.0cm}%
\includegraphics[width=10.3cm]{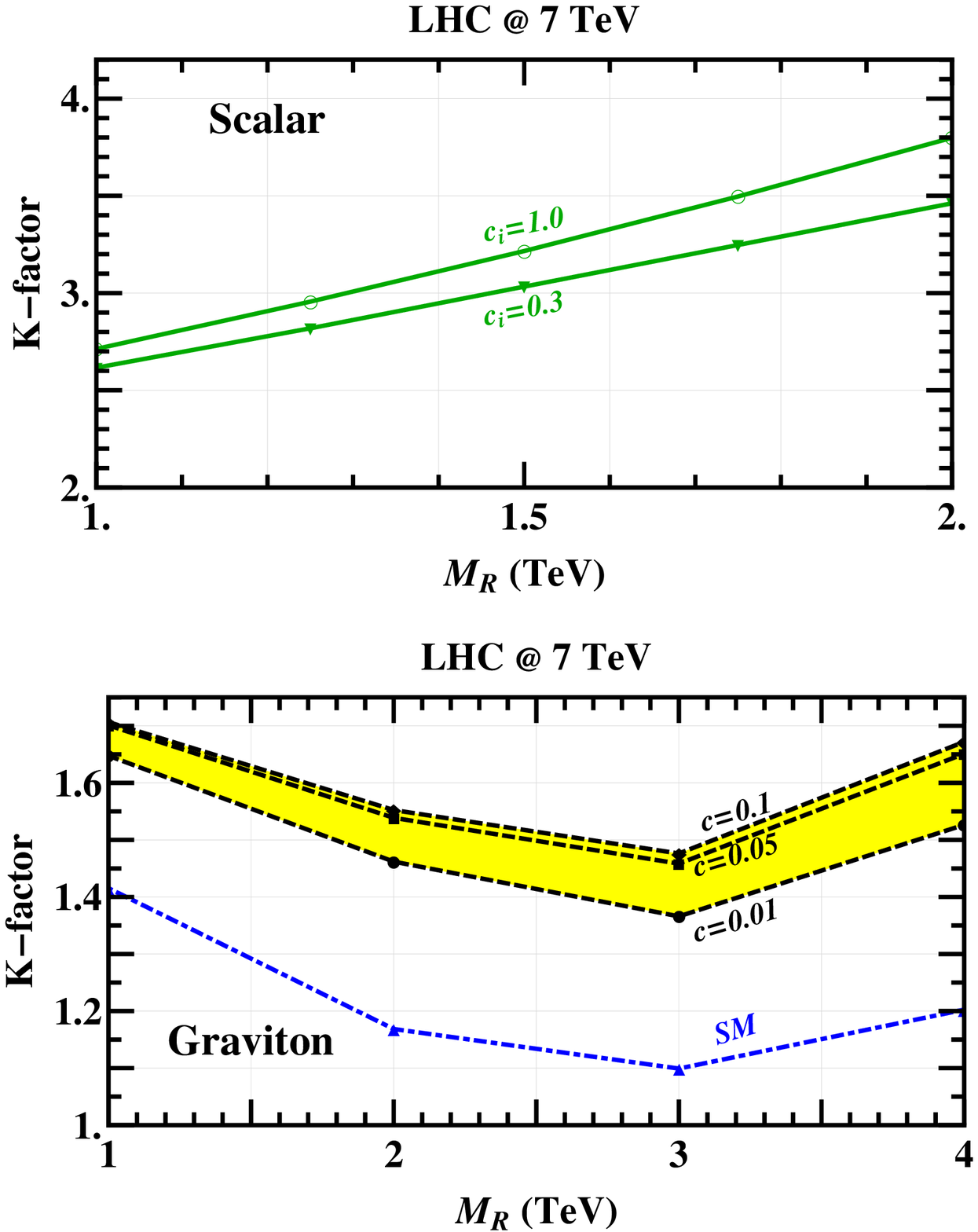}}
\caption{\label{fig-Kfac-MR} K-factor {\em vs} $M_R$ for 14~TeV (left
panel) and 7~TeV (right panel) LHC. }
\end{figure}

\begin{figure}[ptbh!] 
\vspace*{0.5cm}
%
\centerline{ \hspace*{0.8cm}
\includegraphics[width=9.3cm]{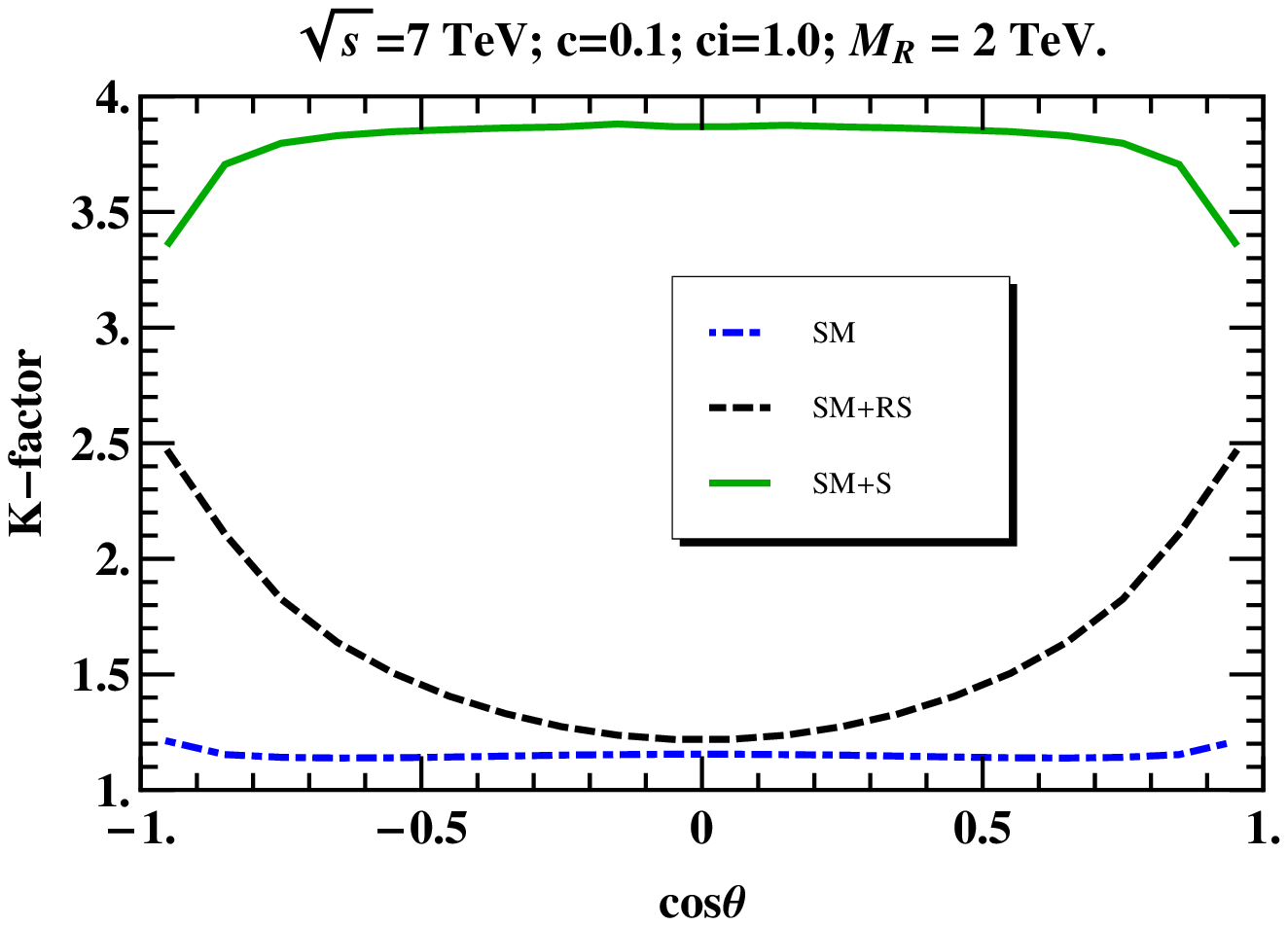}
\hspace*{-1.0cm}%
\includegraphics[width=9.3cm]{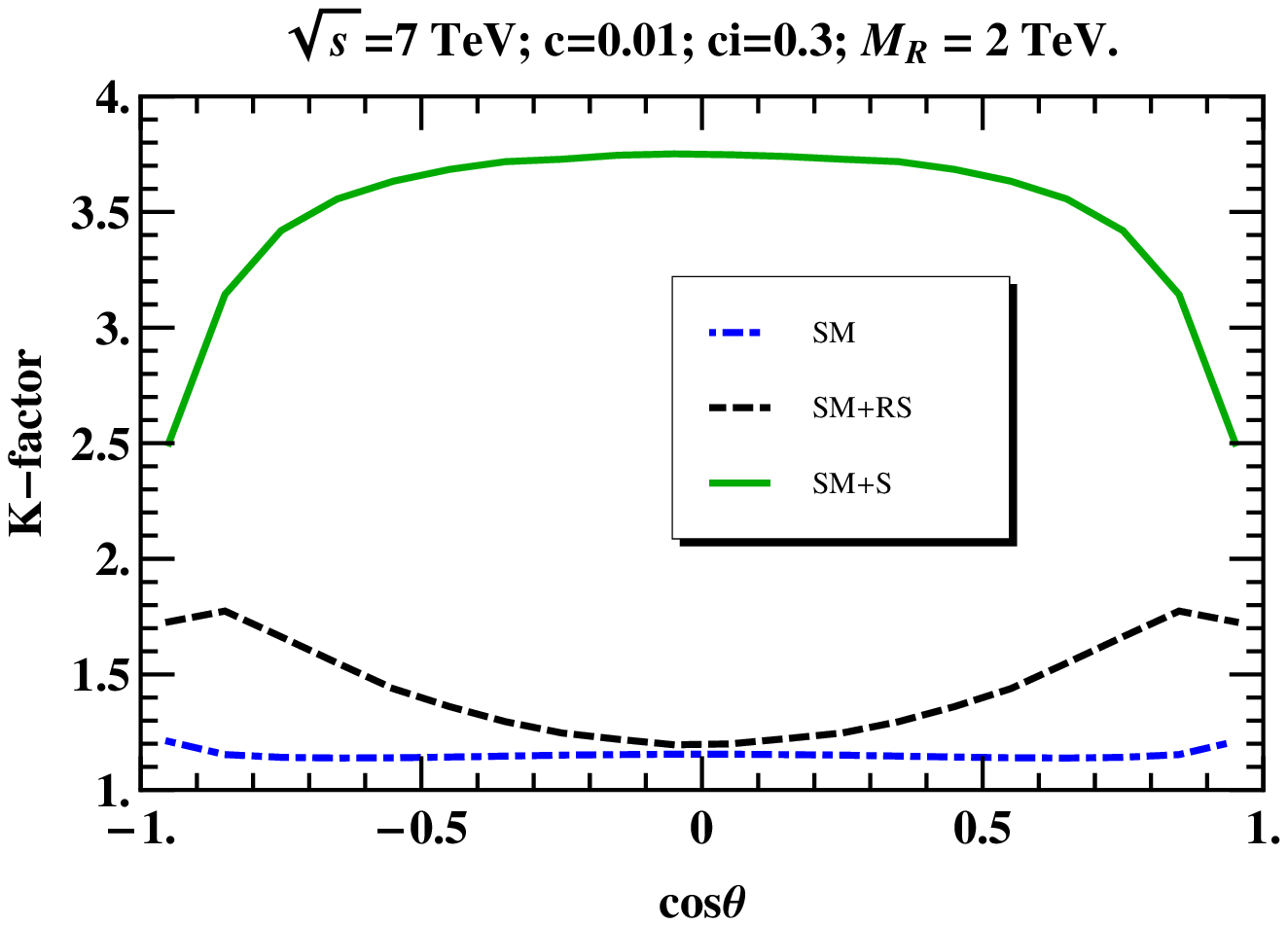}}
\caption{\label{fig-Kfac-z-7tev} K-factor {\em vs} $cos \theta$
for the 7~TeV LHC. }
\end{figure}

\begin{figure}[ptbh!] 
\vspace*{0.5cm} \centerline{ \hspace*{0.8cm}
\includegraphics[width=9.3cm]{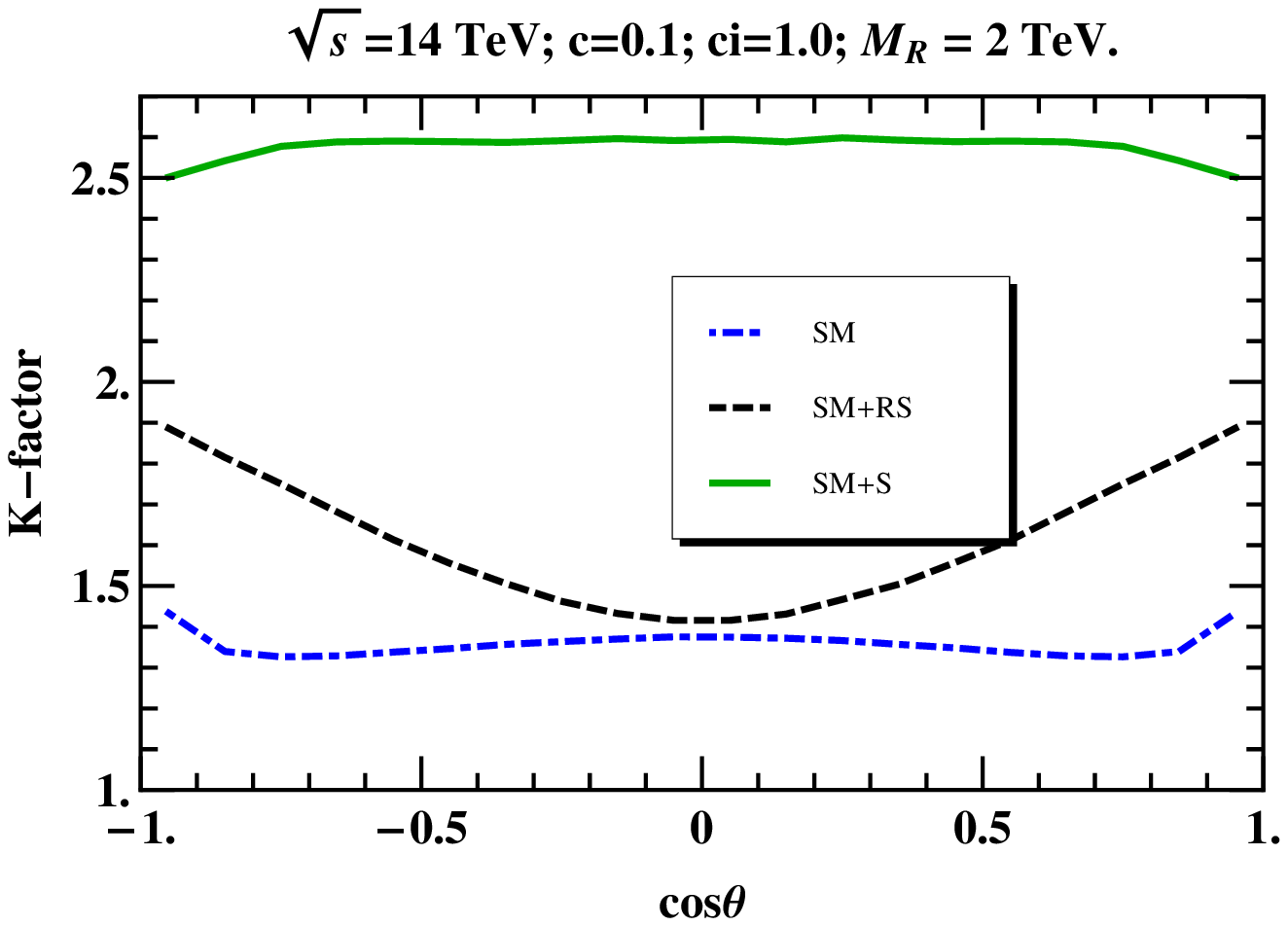}
\hspace*{-1.0cm}%
\includegraphics[width=9.3cm]{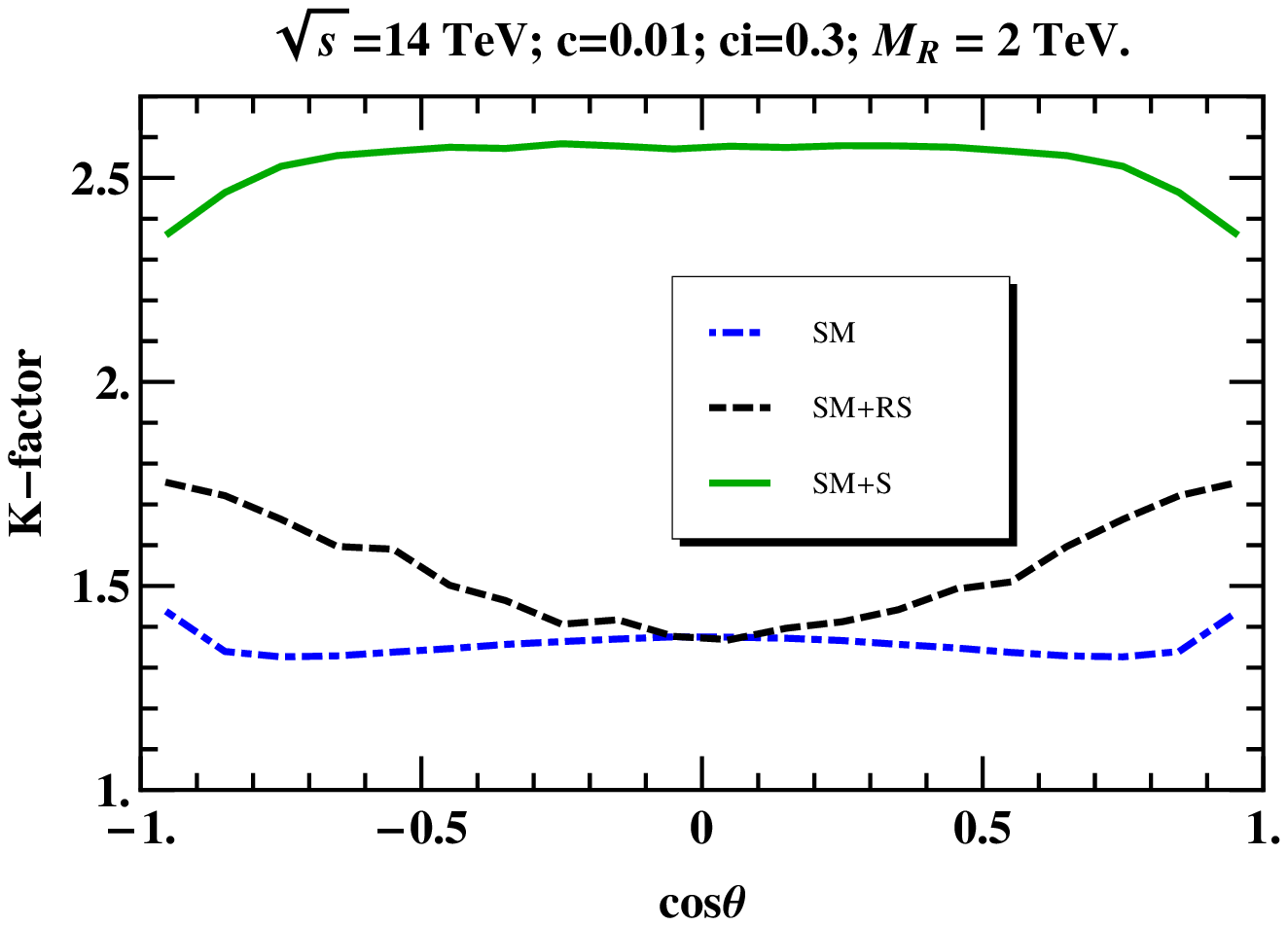}}
\centerline{ \hspace*{0.8cm}
\includegraphics[width=9.3cm]{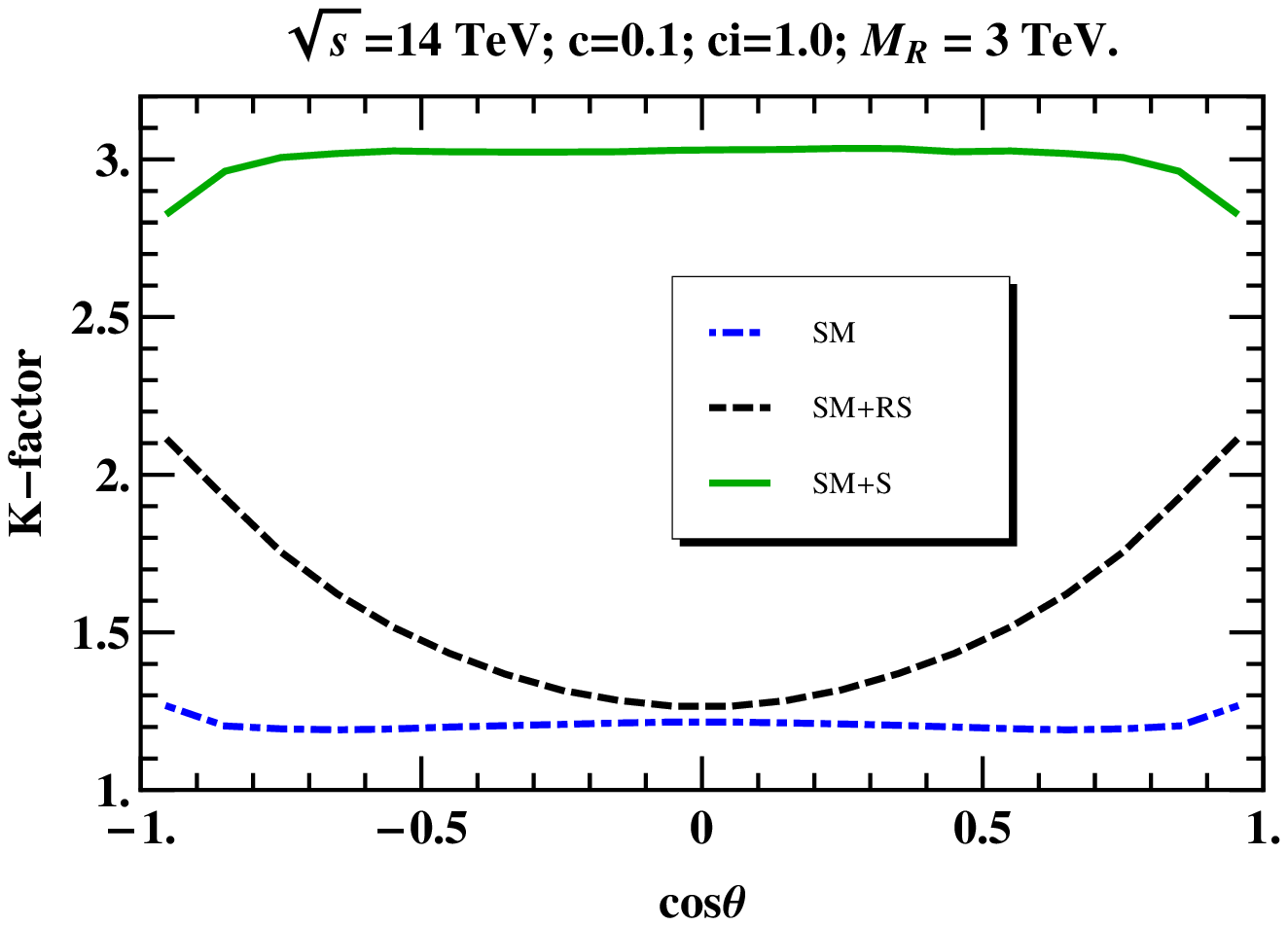}
\hspace*{-1.0cm}%
\includegraphics[width=9.3cm]{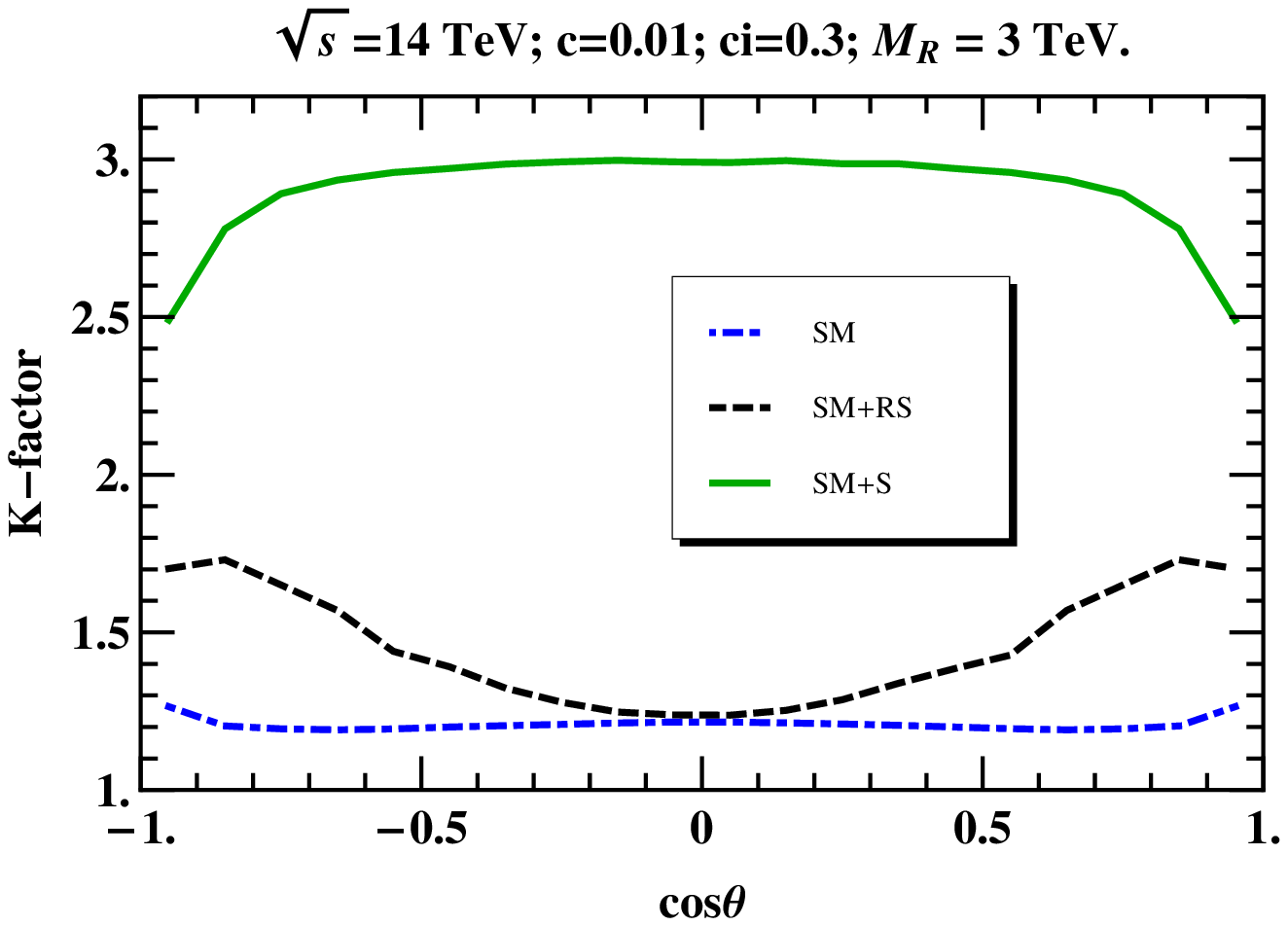}}
\centerline{ \hspace*{0.8cm}
\includegraphics[width=9.3cm]{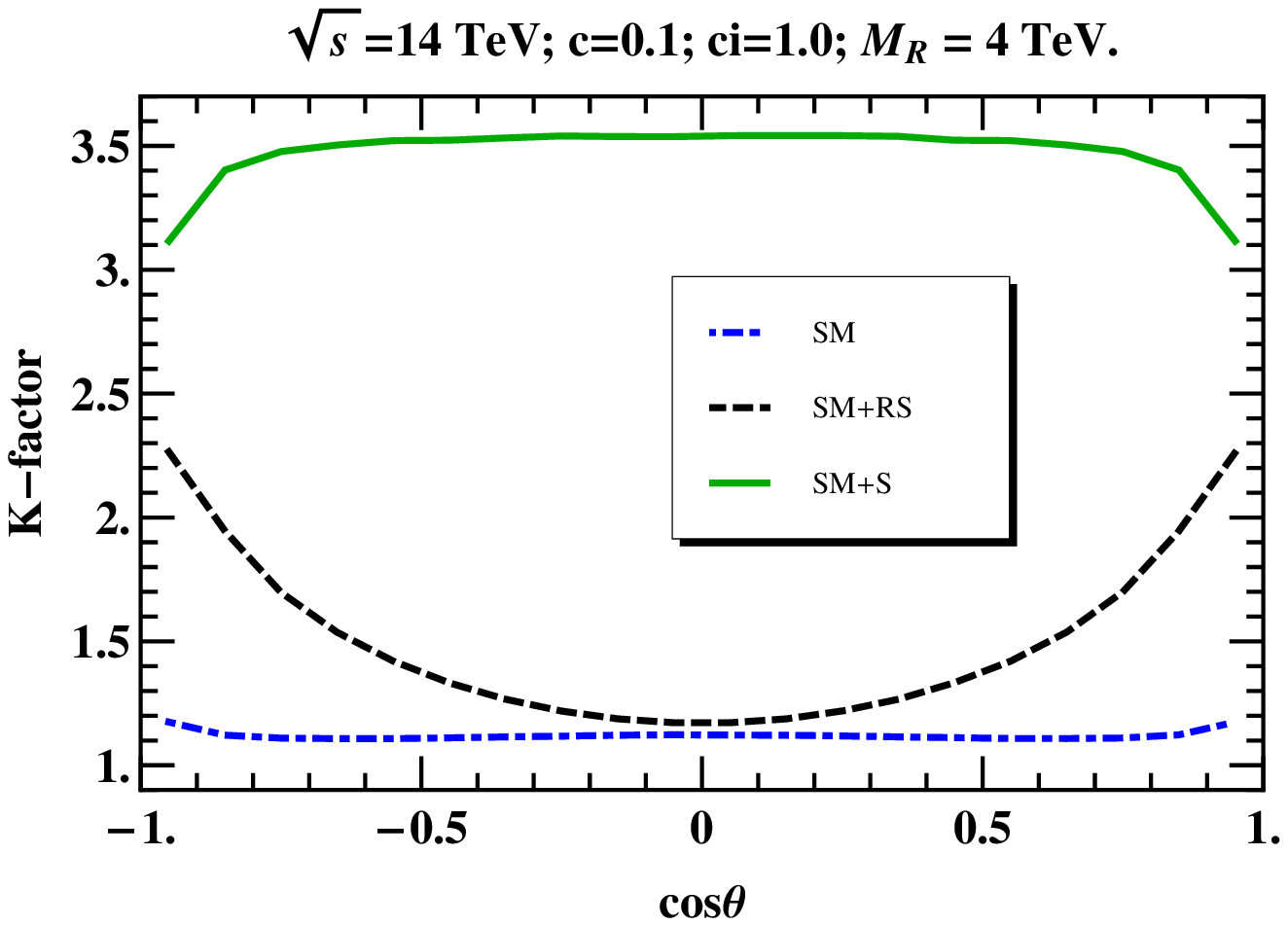}
\hspace*{-1.0cm}%
\includegraphics[width=9.3cm]{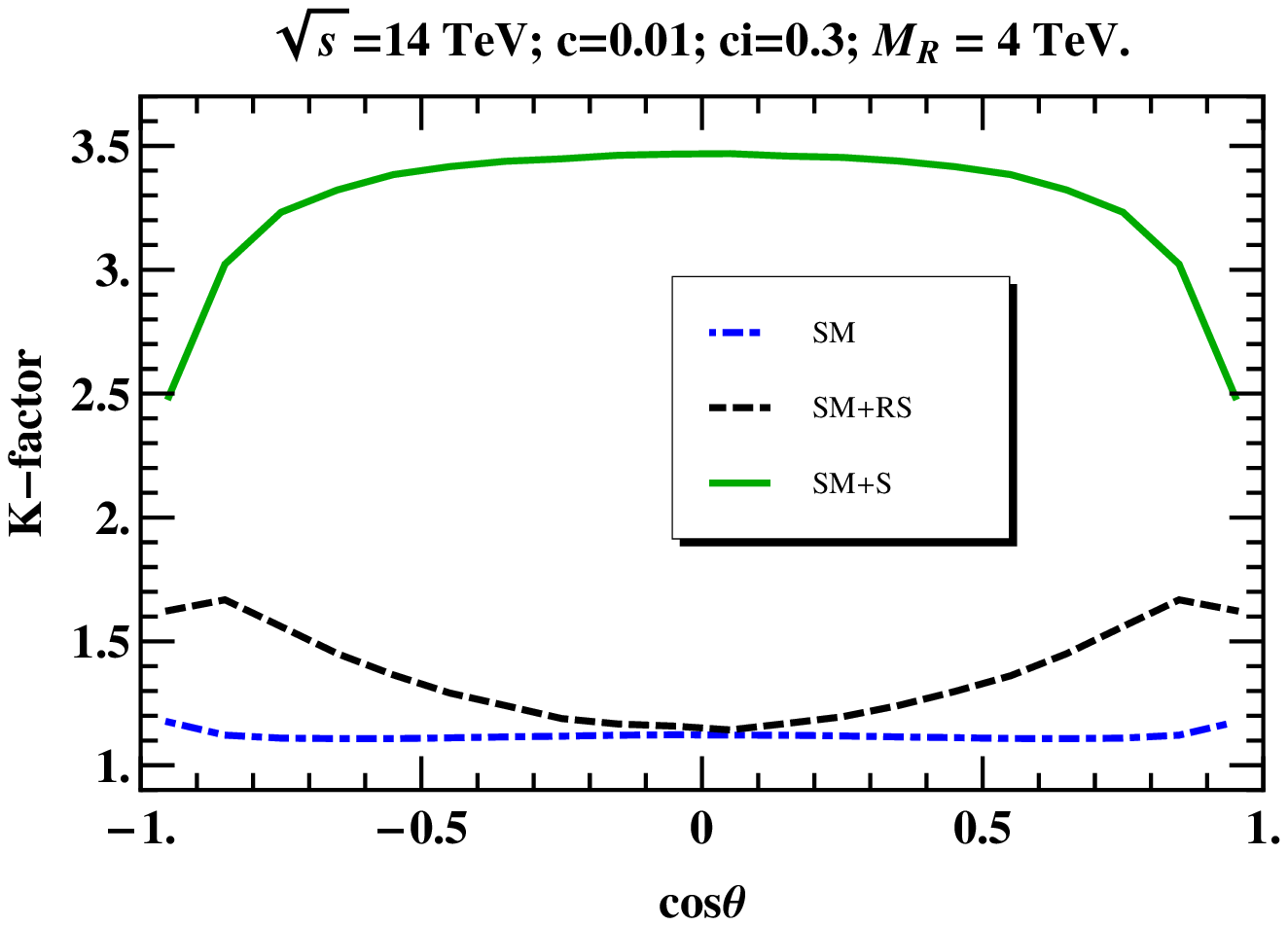}}
\caption{\label{fig-Kfac-z-14tev} K-factor {\em vs} $cos \theta$
for the 14~TeV LHC. }
\end{figure}
%
\par
The K-factors represent quantitatively the magnitude of the NLO QCD corrections, and are
defined as the ratio of the NLO cross section to the corresponding LO one as follows:
\begin{equation}
{\rm K}=\frac{\sigma^{\rm NLO}_{\rm SM} + 
\sigma^{\rm NLO}_{\rm NP}} {\sigma^{\rm LO}_{\rm SM}+
\sigma^{\rm LO}_{\rm NP}} \, .
\label{Kfactor}
\end{equation}
Here, the subscript $NP$ refers to the New Physics contribution (extra dimension or
scalar exchange) and its interference with the SM to LO or NLO as the case may be, while
the subscript $SM$ refers to only the SM contribution.  The ratio could be of total cross
sections as is the case of Fig.\ \ref{fig-Kfac-MR}, or of angular distributions as is the case
of Figs.~\ref{fig-Kfac-z-7tev} and \ref{fig-Kfac-z-14tev}.  Including higher order QCD
corrections to an observable at the hadron collider reduces the dependence on the
factorization and renormalization scales, further since the K-factor itself could be large,
it is essential to include these corrections.
\par
The dependence of the K-factors on the coupling constant $c$ in the RS case, exhibited in the lower
panel of Fig.\ \ref{fig-Kfac-MR},
can be understood by the fact that for low values of
$c \sim 0.01$ the NP
interference with the SM could be of the same order as the pure NP part. For values of
$c>0.05$, the SM contribution is negligibly smaller than the pure NP part and can be
neglected both in the numerator and in the denominator of Eq.\ (\ref{Kfactor}), which in
this case is determined by the NP solely so that the dependence from $c$ cancels.
It is interesting to notice from the upper panel
of Fig.\ \ref{fig-Kfac-MR} that the K-factors in the scalar case are larger than those in the RS case: the
NLO corrections have enhanced the cross sections but,
as shown in Figs.\ \ref{fig-Kfac-z-7tev} and
\ref{fig-Kfac-z-14tev}, they did not noticeably change
the shape of the angular distributions from the flat behavior of the pure scalar
particle exchange cross section
in Eq.\ (\ref{dsigmaggs}).
\par
Also, one may remark that in the example discussed here the RS
graviton couples to quarks and gluons with equal coupling strength
whereas, in the model of Ref.~\cite{Barbieri:2010nc} we have
adopted, the scalar particle couples mainly to gluons (couplings
to quarks are identically zero in the limit of vanishing quark masses). Furthermore, the
production of a scalar particle in the gluon-gluon fusion
subprocess is {\it qualitatively} equivalent to the Higgs boson production in the limit of infinite top quark mass: it is
well-known that K-factors (due to NLO QCD corrections) for the
Higgs production process at hadron colliders are very high and can
easily be greater than 2.0 in the light Higgs mass region
\cite{Graudenz:1992pv, Djouadi:1991tka, Dawson:1990zj}.  Hence, a
similar pattern of K-factors can be expected in the case of a
scalar production (followed by decay to photons) in the model
considered here, see  Figs.\ \ref{fig-Kfac-MR},
\ref{fig-Kfac-z-7tev}, and \ref{fig-Kfac-z-14tev}.

\section{Angular analysis and RS graviton identification}
\begin{figure}[tbh!] 
\vspace*{0.5cm} \centerline{ \hspace*{0.5cm}
\includegraphics[width=9.0cm]{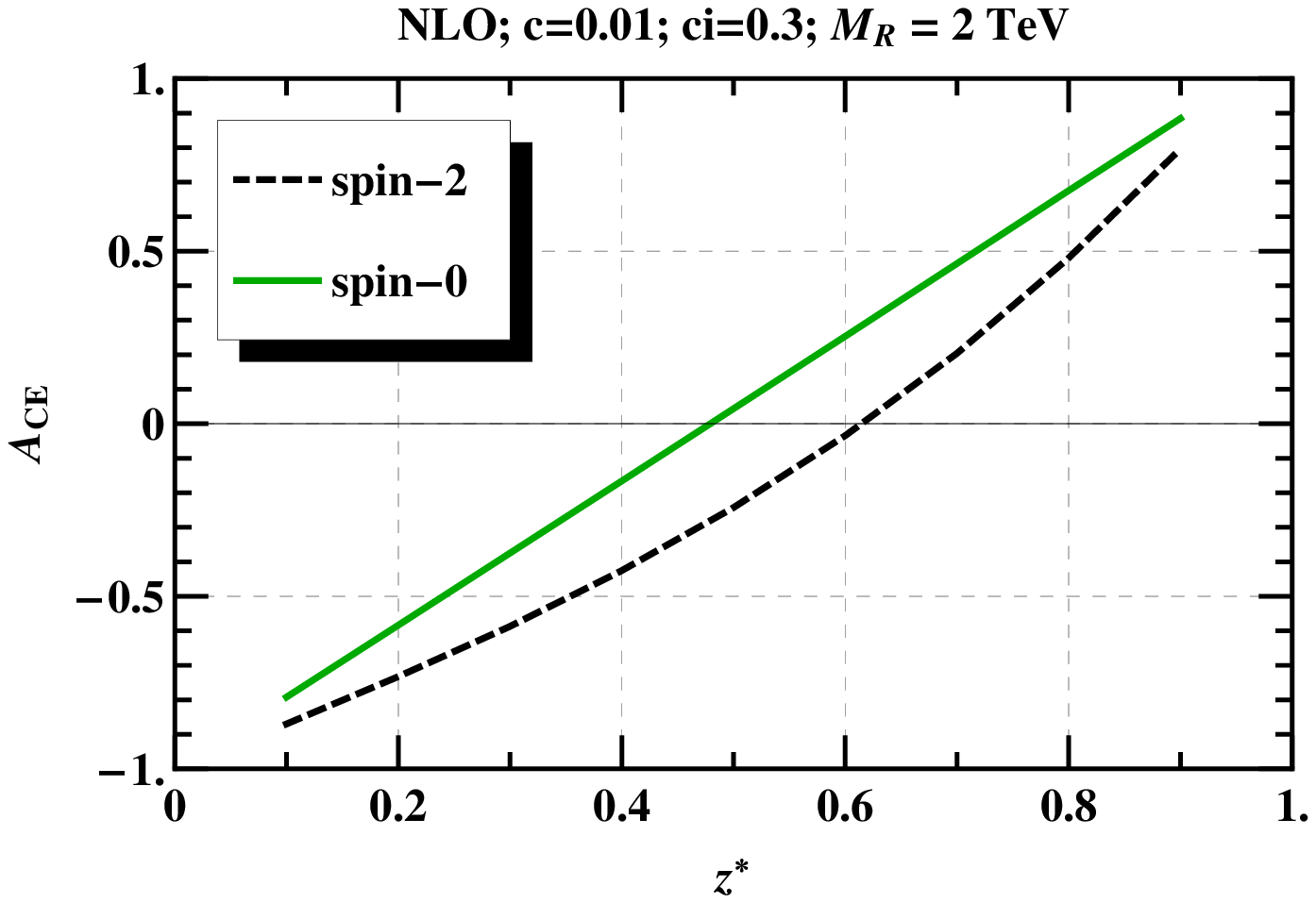}
\hspace*{-1.0cm}%
\includegraphics[width=9.0cm]{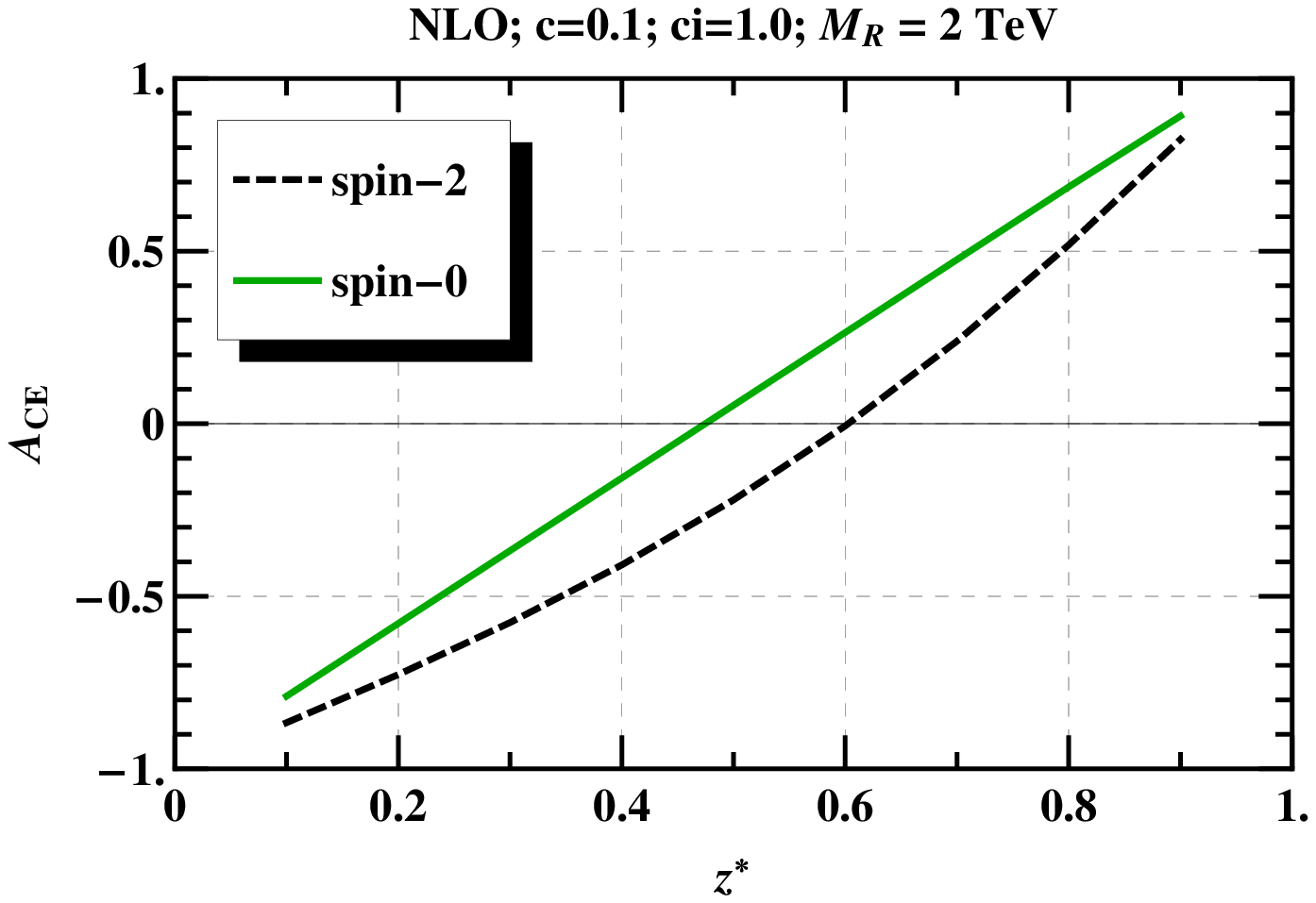}}
\centerline{ \hspace*{0.5cm}
\includegraphics[width=9.0cm]{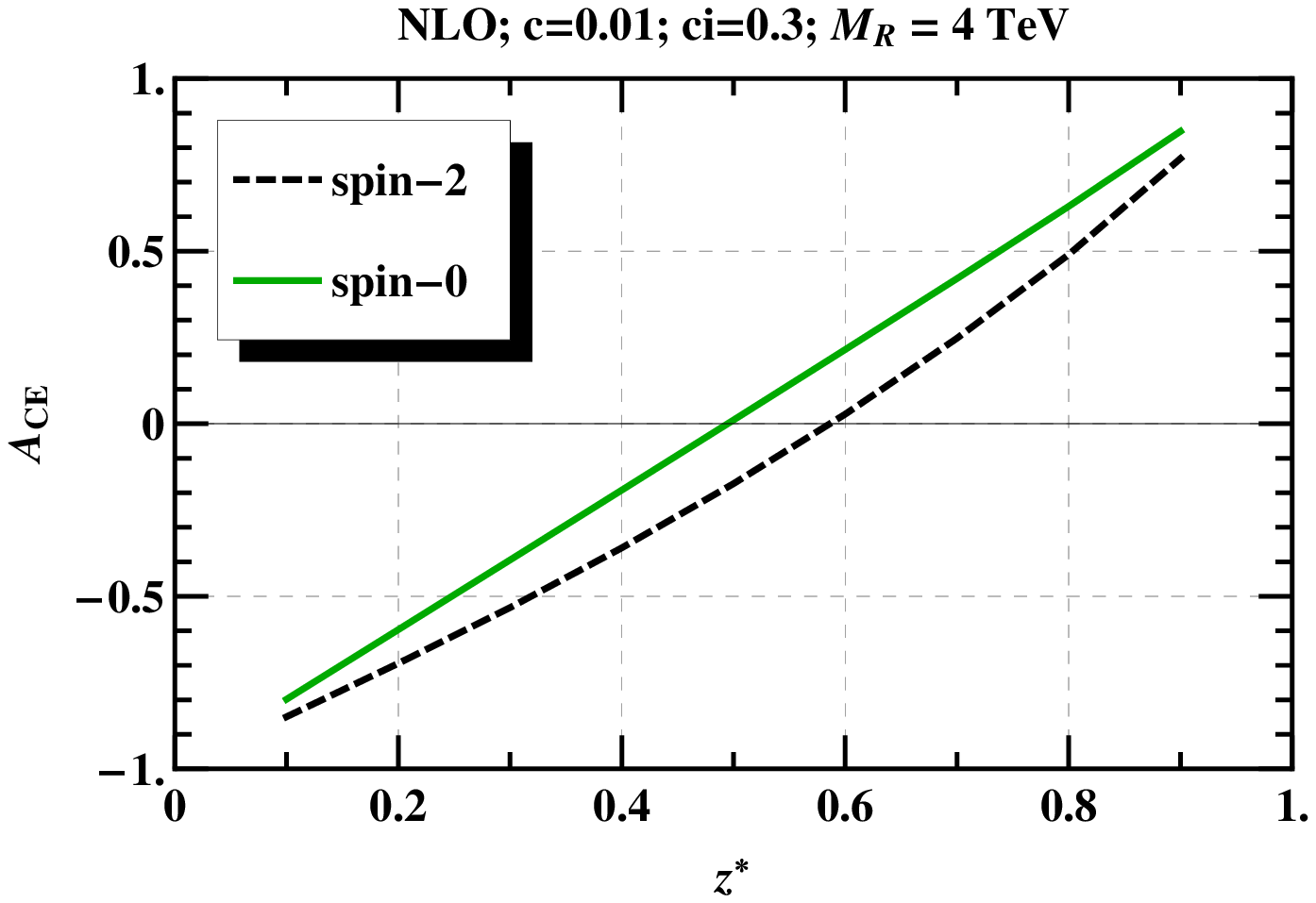}
\hspace*{-1.0cm}%
\includegraphics[width=9.0cm]{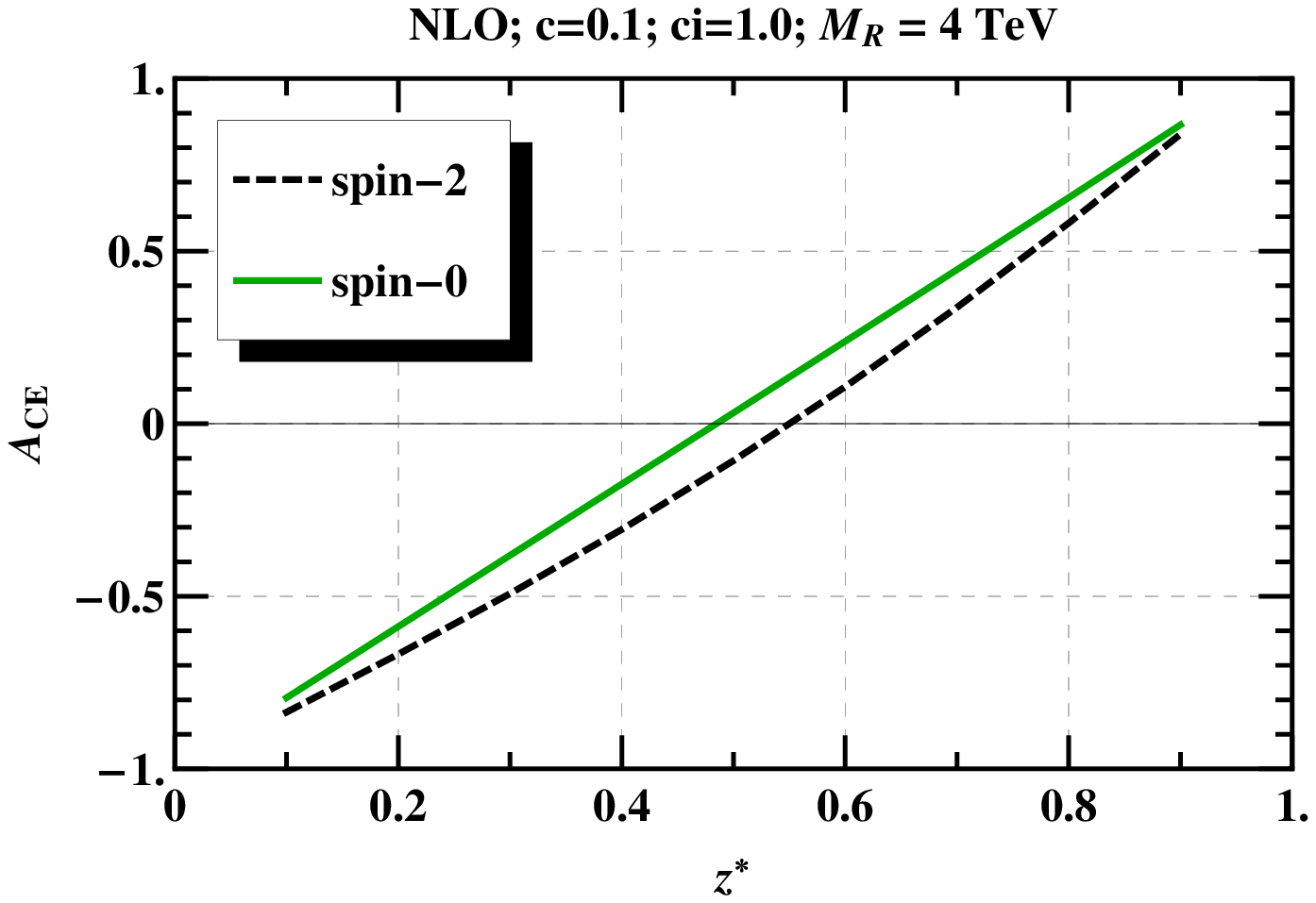}
} 
\caption{\label{fig-ace-zst} NLO $A_{\rm CE}$ asymmetry vs $z^{*}$ for the inclusive process (\ref{proc}) at the 14 TeV LHC for the RS model with graviton coupling constant 
$c=0.01$ $(0.1)$, and the scalar resonance model with $c_3=0.3$, $c_i=0.3$ ($c_3=0.3$, $c_i=1.0$). 
$M_R=2$ TeV: top-left panel (top-right panel).
Bottom panels: same as for top panels, but with $M_R=4$ TeV.}
\end{figure}

The $A_{\rm CE}$-based angular analysis will essentially proceed
as follows. The first step will be the determination of
graviton-scalar ``confusion regions'', namely, of the subdomains
in respective discovery signature spaces of coupling constants and
masses where, for $M_R=M_G=M_S$, the two models predict equal
numbers of resonance signal events, $N_S$, hence are not directly
distinguishable on a statistical basis. In such confusion regions,
one then can try to discriminate the models from one another by
means of the different values of the $A_{\rm CE}$ asymmetry
generated by the respective photon angular distributions. The
representation of $A_{\rm CE}$ {\em vs} $z^*$ predicted at NLO for the
two models with diphoton resonance masses of 2 and 4 TeV, at the
14 TeV LHC, is shown in Fig.~\ref{fig-ace-zst}.

%
\par
One can start from the assumption that an observed peak at $M=M_R$
is due to the lightest spin-2 RS graviton (thus, that $M_R=M_G$),
and define a ``distance'' from the scalar-exchange model
hypothesis as:
\begin{equation}
\Delta A_{\rm CE}^{G}=
A_{\rm CE}^{G}-A_{\rm CE}^{S}.
\label{deltaGS}
\end{equation}
An indication of the domain in the ($M_G,c$) RS parameter plane,
where the competitor spin-0 hypothesis giving same number of
resonant events for $M_R=M_S=M_G$ can be excluded by the
starting RS graviton hypothesis, can be obtained from a
simple-minded $\chi^2$-like numerical procedure, similar to that used in Ref.~\cite{Osland:2008sy}. The comparison of the
deviations (\ref{deltaGS}) to the statistical uncertainty $\delta
A_{\rm CE}^G$ pertinent to the RS model,
suggests the following criterion for spin-0
exclusion:
\begin{equation}
\chi^2\equiv
\vert\Delta A_{\rm CE}^{G}/\delta A_{\rm CE}^{G}\vert^2
> \chi^2_{\rm CL}.
\label{chisquare}
\end{equation}
Eq.~(\ref{chisquare}) shows the definition of the $\chi^2$, and
$\chi^2_{\rm CL}$ specifies a desired scalar-exchange exclusion confidence level
(for example, $\chi^2_{\rm CL}=3.84$ for 95\% CL). With
$\Delta A_{\rm CE}^{G}$ calculated in terms of $M_R$ and
model coupling constants, this condition will define the
domain in the confusion-regions of model parameters
where the RS spin-2 hypothesis can be discriminated from the scalar exchange.
With
$(A_\text{CE}^G)^2$ much smaller than unity for values of $z^*$ around 0.5, we have to a good approximation:
\begin{equation} \label{Eq:stat}
\delta A_{\text{CE}}^{G}
=\sqrt{\frac{1-{(A_\text{CE}^G)}^2}{N_{S,\rm{min}}}} \approx
\sqrt{\frac{1} {N_{S,\rm{min}}}},
\end{equation}
where $N_{S,\rm min}$ will be the minimum number of RS resonance events
needed to satisfy the criterion (\ref{chisquare}), hence
to exclude the spin-0 exchange model
with same $M_R=M_G=M_S$ in the confusion-region of the parameters.
The knowledge of $N_{S,\rm min}$ determines, in turn, the RS resonance identification subdomain in the ($M_G,c$) parameter plane.

\begin{figure}[t!] 
\vspace*{-0.5cm}
\centerline{ 
\includegraphics[width=8.0cm,angle=0]{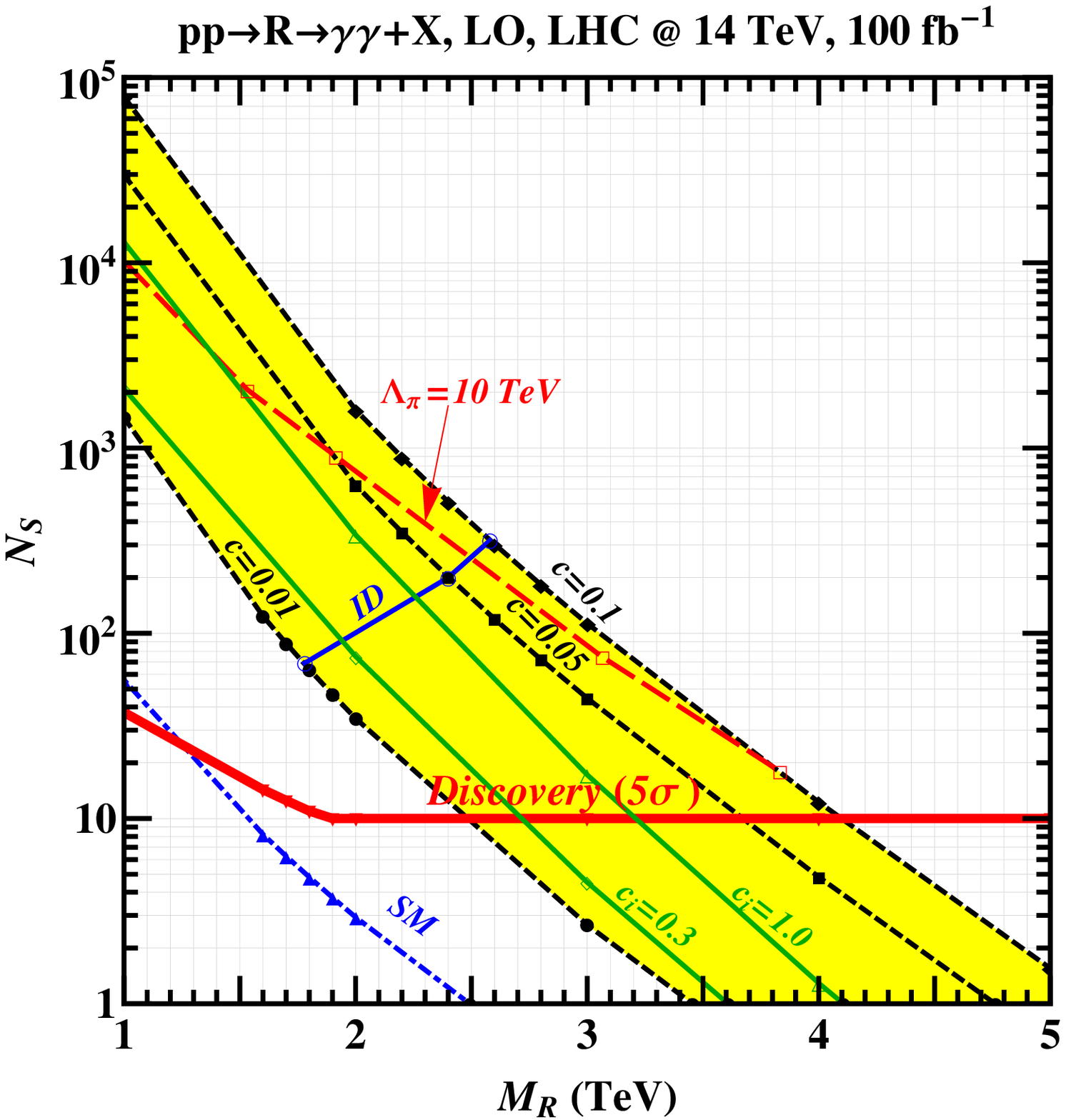}
\includegraphics[width=8.0cm,angle=0]{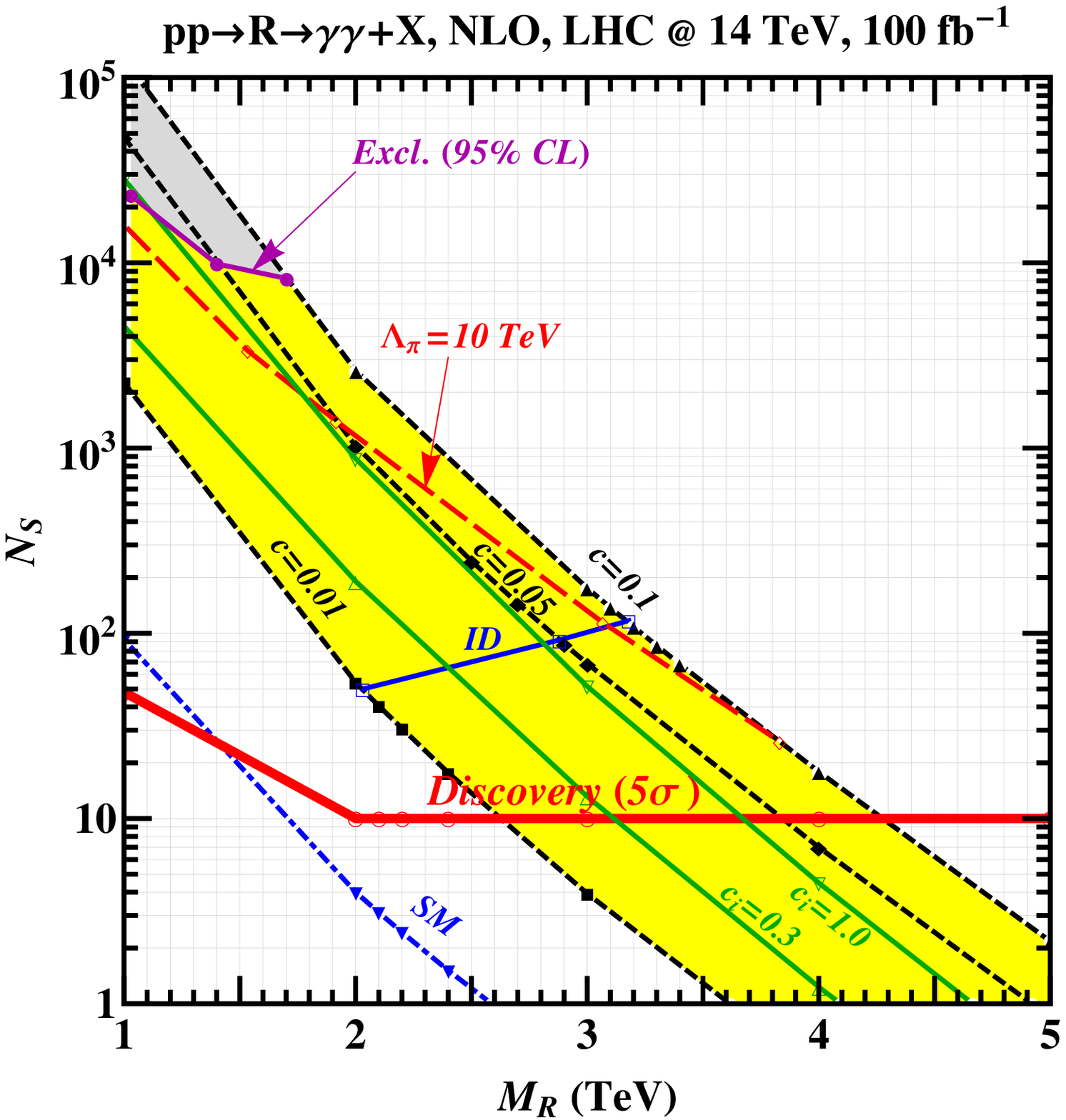}}
\caption{\label{fig-sig-14tev}  Number of resonance (signal)
events $N_S$ {\em vs} $M_{R}$ ($R=G,\,S$) at the LHC with
$\sqrt{s}=14$~TeV and $\Lumint=100~\text{fb}^{-1}$ for the process
$pp\to G\to \gamma\gamma+X$ at LO (left panel) and NLO (right
panel) QCD. The yellow area corresponds to the KK graviton signature
space for $0.01\le c \le 0.1$; the signature space for the scalar $S$ resonance is represented by the two specific cases, $c_3=c_i=0.3$ and $c_3=0.3$, $c_i=1.0$. }
\end{figure}
\begin{figure}[hb!] 
\vspace*{-0.5cm}
\centerline{ 
\includegraphics[width=8.0cm,angle=0]{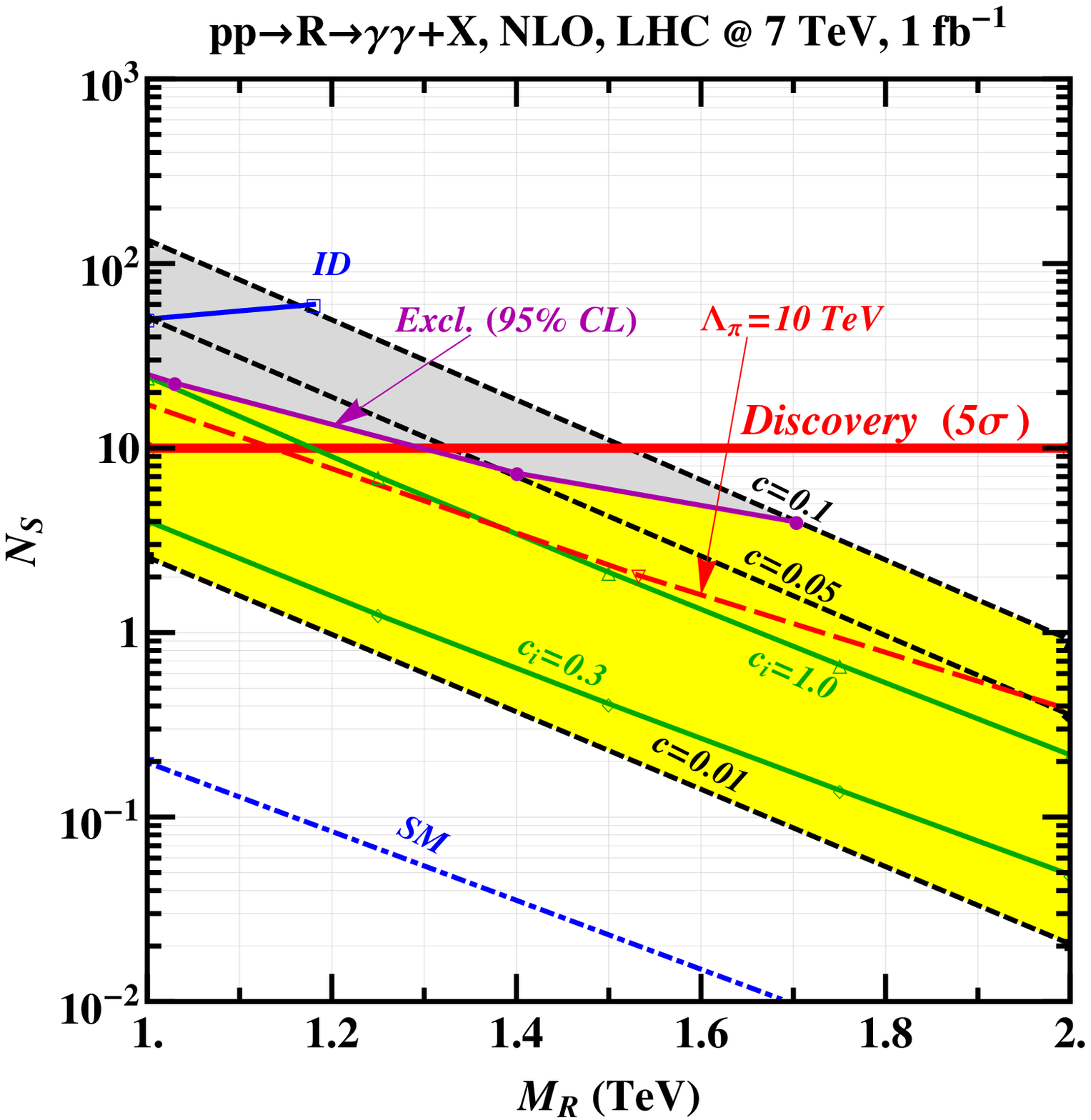}
\includegraphics[width=8.0cm,angle=0]{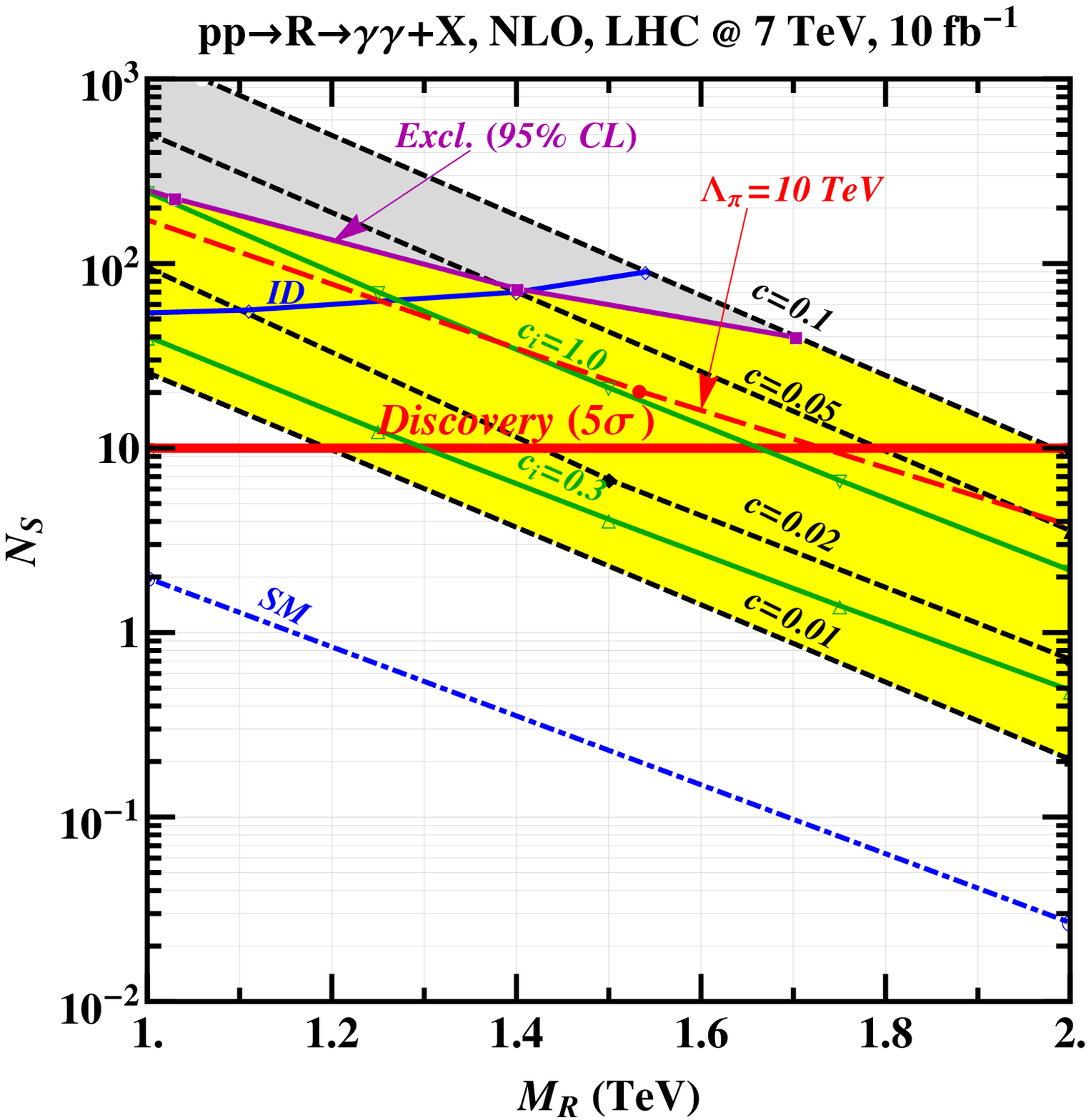}}
\caption{\label{fig-sig-7tev}  Similar to
Fig.~\ref{fig-sig-14tev}, but for NLO QCD at $\sqrt{s}=$7~TeV and
$\Lumint=$1~fb$^{-1}$ (left panel) and $\Lumint=$10~fb$^{-1}$
(right panel).}
\end{figure}
\par
We apply this procedure to the design LHC running conditions
$\sqrt s=14$ TeV and time-integrated luminosity ${\cal L}_{\rm
int}=100$ ${\rm fb}^{-1}$. Figure~\ref{fig-sig-14tev} shows the RS
graviton signature domain for $0.01\leq c\leq 0.1$ at LO (left
panel) and to NLO (right panel) in QCD. The red line labelled as ``Discovery'' indicates the minimum number of events statistically needed for the RS KK resonance discovery in process (\ref{proc}) at the
5-$\sigma$ level, as anticipated in Sec.~II. The
domain between the lines labelled as $c=0.01$ and $c=0.1$ represents the number of events for RS KK graviton production followed by decay into a photon pair, theoretically evaluated as described in the previous section, for different values of $M_G$. The scalar resonance signature space is also
included in this figure, at the same LO and NLO,
for simplicity by the
representative lines labelled as $c_3=c_i=0.3$ and $c_3=0.3,\
c_i=1.0$ ($i\neq 3$), respectively. One can see that, for these
values of the scalar coupling constants, there is a finite confusion
region where, at given mass $M_R=M_G=M_S$, the numbers of
predicted resonance signal events can be equal. Actually, such a
confusion region might easily be extended to almost
completely overlap with the full RS signature space by partially weakening the condition $\Gamma_S\leq\Delta M$. This condition is, anyway, to be understood in a qualitative sense, so that more numerical freedom to the scalar coupling constants of Eq.~(\ref{scalar}) might be allowed.
Indeed, if the width $\Gamma_S$ turned out to be larger than
$\Delta M$ in Eq.~(\ref{deltam}), the analysis proposed here
should still be viable, and could in this case  discriminate a narrow KK graviton
vs the scalar resonances, both by the angular observable $A_{\rm CE}$ and by the size of the widths themselves. Also, we can remark that, as
relying just on specific angular distributions, the kind of
analysis proposed here should be applicable more generally, to the identification of the RS graviton excitation from
different scalar exchange models than studied here.
\par
The line labelled as ``ID'' in Fig.\ \ref{fig-sig-14tev}
essentially represents the solution of Eq.~(\ref{chisquare})
relating $M_G$ to $c$, i.e., it is the
minimum number of events needed for a discovered RS graviton
resonance to be identified at 95\% CL against the scalar particle exchange hypothesis, according to the $A_{\rm CE}$-based angular analysis.
The differences between the left and right panels clearly show the need to account for the large NLO QCD effects
in the theoretical description of the resonant diphoton inclusive production at LHC.
\par
In practice, as one can read from the right, next-to-leading order panel in Fig.\ \ref{fig-sig-14tev}, if an RS graviton excitation will be discovered in diphoton events at the
14 TeV LHC with 100 ${\rm fb}^{-1}$ luminosity, its spin
may be identified for $M_G$ up to, roughly, 2--3 TeV
for $c=0.01 - 0.1$ if the number of observed RS resonance signal events will be larger or equal to those indicated by the line ``ID'': namely, $N_{S,min}\gtrsim 50 - 120$. Of course, these indications follow from the
criterion (\ref{chisquare}) outlined above, hence rely on
statistical arguments and theoretical calculations of the relevant cross sections, in particular one may notice in this regard that the SM background turns out to be completely negligible
with respect to the signal. The detailed assessment of the ``experimental'' backgrounds to the resonance discovery
in process (\ref{proc}), and of the related systematic uncertainties, is out the scope of this paper, our purpose here is to just compare (and discriminate) two different theoretical explanations for same resonance mass events, once observed, on the basis of NLO calculations in QCD.
\par
The grey area in the right panel of Fig.\ \ref{fig-sig-14tev} represents the 95\% CL
exclusion, where the RS resonance should not be observable, if we account for the
lower limits on $M_G$ vs $c$ derived from the dilepton production analyses recently presented in Refs.~\cite{atlas2} and \cite{cms2}. Thus, the range in
$M_R$ of interest would start, in view of the new LHC results, from $M_G=1.6-1.7$
TeV for $c=0.1$ and $M_G=0.7$ TeV for $c=0.01$.
\par
Also drawn in Fig.\ \ref{fig-sig-14tev} is the line
$\Lambda_\pi=10$ TeV: the ``theoretical'' condition
$\Lambda_\pi<10$ TeV mentioned in Sec. III., if enforced
literally, would dramatically constrain the RS discovery
domain in the plane $(M_G,c)$ to the events located above this line. However, also this condition should be understood in a qualitative sense, as is the case, in principle, of
the assumed range of values for $c$.
\par
To make a contact with the current LHC running conditions,
we repeat the RS identification procedure outlined above for the 7 TeV case, with
time-integrated luminosities ${\cal L}_{\rm int}=1$
and 10 ${\rm fb}^{-1}$.
The analogue of Fig.~\ref{fig-sig-14tev}, but this time for the NLO calculations only, is represented in
Fig.~\ref{fig-sig-7tev}, with the same significance of the symbols.
The interpretation of the left and right panels in this figure is also completely analogous. For example,
if a RS resonance were discovered in diphoton events
at 1 ${\rm fb}^{-1}$, its spin-2 might be
identified, to 95\% CL, up to $M_G=1.2$ TeV for $c=0.1$,
provided the collected signal
were about $60$ events; and for
10 ${\rm fb}^{-1}$, identification would be possible up
to $M_G=1.5$ TeV for $c=0.1$ with a collected signal
of, say, about $100$ events.
\par
However, the situation is drastically modified by the 95\% CL exclusion limits from
the dilepton analysis of Refs.~\cite{atlas2,cms2} at 1.2 ${\rm fb}^{-1}$, reported in Sec.~III and represented by the grey areas in both panels of Fig.~\ref{fig-sig-7tev}. The experimental limit
$M_G> 1.6-1.7$ TeV for $c=0.1$ is not quite inconsistent with the left panel of Fig.~\ref{fig-sig-7tev}, which
shows that for these values the theoretically
predicted statistics for RS events falls below that
needed for 5-$\sigma$ discovery.
The exclusion range starts from
the ``low'' values $c=0.01$ and $M_G=0.7$ TeV, so that
the left panel of this figure shows that, at the present
stage, in principle there may be a little corner left
available for discovery, roughly $M_G$ between 0.7 and
1.3 TeV and small $c<0.05$. On the other hand, this panel
clearly indicates that there is no room for
RS graviton identification at 1 ${\rm fb}^{-1}$ luminosity,
at least with the angular analysis presented here.
Moving to the 10 ${\rm fb}^{-1}$ case, the right panel of
Fig.\ \ref{fig-sig-7tev} shows that, with the current LHC limits, discovery might still be possible up to about $M_G=2$ TeV with $c=0.1$, but the identification would be allowed only for $M_G=0.7-1.4$ TeV with $c$ not larger than, say, $0.05$.

\section{Final considerations}
In the previous sections, we have discussed the features of different-spin
$s$-channel exchanges in the inclusive diphoton production process
(\ref{proc}) at the LHC. Specifically, we have considered the hypothesis
of the spin-2 RS graviton excitation exchange as the source of an  eventually
discovered peak in the diphoton invariant mass, and compared it to the
interpretation of the same peak as a  spin-0 scalar exchange exemplified by
model~\cite {Barbieri:2010nc}. The aim has been of determining quantitatively
the domain in the RS graviton mass $M_G$ and coupling constant
$c=k/{\overline M}_{\rm Pl}$, where the former hypothesis can be identified
against the latter (the so-called identification reach).
\par
Clearly, to this purpose, in the situation of equal number of peak events
from the two models, the information from the distinctive photon angular
distributions is needed. The relevant cross sections, differential and
angular integrated, at next-to-leading order in QCD have been introduced
for both the RS and the scalar-exchange model. The comparison with the
leading order calculations shows that the NLO effects are substantial,
and non-negligible for this analysis. As shown in Sec.~IV, in the TeV
resonance mass range of interest here and at the considered LHC energies,
K-factors turn out to be large, of the order of 1.5 or so for the RS
exchange, and even larger, of the order of 3 or so, for the scalar
exchange model. Moreover, while K-factors exhibit an angular dependence
in the case of the RS model, they do not noticeably alter the flat shape
of the angular distribution for the pure scalar resonance exchange derived
at leading order in QCD.
%
\par
The angular analysis to discriminate the RS from the scalar model has been based on the center-edge asymmetry 
$A_{\rm CE}$, also estimated at NLO in QCD, 
for both the 14 TeV and the 7 TeV LHC. The numerical results for the
predicted total number of resonant events and the minimum number of events
for RS identification, obtained by the simple statistical arguments outlined
in the previous section, are presented in Figs.~\ref{fig-sig-14tev} and
\ref{fig-sig-7tev}, and summarized in Table~\ref{table:dis-id}.
\begin{table}[htb]
\begin{center}
\begin{tabular}
[c]{|c|c|c|c|c|}\hline
 $\sqrt{s}, \Lumint$ & \multicolumn{2}{|c}{\,\, 14 TeV, 100 fb$^{-1}$} & \multicolumn{2}{|c|}{\,\, 7 TeV, 10
fb$^{-1}$} \\\hline\hline
$k/{\overline M}_{\rm Pl}$ & \,\,\,\, Dis \,\, & ID & \,\, Dis
\,\, & ID \\\hline
0.01 & 2.6 & 2.0 & 1.2 & $\sim$0.9 \\\hline
%
%
0.05 & 3.8 & 2.9 & 1.8 & 1.4 \\\hline
0.1 & 4.2 & 3.2 & 2.0 & Excl. \\\hline
\end{tabular}
\caption{Theoretical discovery ($5\sigma$-level) and 95\% CL
identification limits (NLO) on $M_G$ in TeV. \label{table:dis-id}}
\end{center}
\end{table}
\par
As regards the 7 TeV LHC, the theoretical results obtained above show that,
taking into account recent experimental limits on the RS graviton mass, for
luminosity around 10 ${\rm fb}^{-1}$ there may still be the possibility to
identify the RS graviton excitation, in a rather limited range of $M_G$ and
small $c$. Higher luminosity, and/or larger LHC energy, would be required
to extend this region by the $A_{\rm CE}$ based angular analysis.
\par
Finally, one may observe that the results regarding the RS graviton
identification obtained in the previous section show that the inclusive
diphoton process (\ref{proc}) at the LHC is complementary to the
Drell-Yan dilepton production, in the sense that only the scalar exchange
needs be considered as a competitor, alternative, hypothesis for the
source of resonance events. We have performed the $A_{\rm CE}$ angular
analysis of the diphoton production cross sections for both hypotheses
at NLO in QCD, and estimated numerical results for discovery and
identification. The $A_{\rm CE}$ method has so far been applied to the
dilepton channel at the LO in QCD only. Therefore, it should be
interesting, for a fully exhaustive comparison of the results
achievable from the two channels, to extend the $A_{\rm CE}$-based
angular analysis at NLO also to the dilepton production process.
%
\newpage
\leftline{\bf Acknowledgments}
\par\noindent
This research has been partially supported by funds of the University of
Trieste and by the Abdus Salam ICTP under the TRIL and the Associates
. PM acknowledges the ICTP associateship, which lasted until
the initial stage of this  collaboration. The work of MCK and VR has
been partially supported by the funds of Regional Center for
Accelerator based Particle Physics (RECAPP), Department of Atomic Energy,
Govt.\ of India. We would also like to thank the cluster computing facility
at Harish-Chandra Research Institute, where part of the computational work
for this study was carried out. A conversation with I.\ A.\ Golutvin is
gratefully acknowledged.

\end{document}